\documentstyle[11pt,aaspp4,tighten]{article}

\def\1821{H~1821$+$643}
\def\2352{PKS~2352$-$342}
\def\9104{IRAS~09104$+$4109}
\def\mcool{\.{M}$_{\rm cool}$}
\def\mdot{\.{M}}

\def\myr{M$_{\sun}$/yr}
\def\ROSAT{\sl ROSAT\rm}
\def\rosat{\sl ROSAT\rm}
\def\rcore{r$_{\rm core}$}
\def\tcool{t$_{\rm cool}$}

\def\bgq{B$_{\rm gq}$}
\def\cgsflux{ergs~cm$^{-2}$~s$^{-1}$}
\def\cm3{cm$^{-3}$}
\def\3cm3{10$^{-3}$~cm$^{-3}$}
\def\ctsec{cts~s$^{-1}$}
\def\ein{\sl EINSTEIN\rm}

\def\ksm{km~s$^{-1}$~Mpc$^{-1}$}
\def\kms{km~s$^{-1}$}
\def\hf{h$_{50}$}
\def\ho{H$_{\rm o}$}
\def\qo{q$_{\rm o}$}
\def\h0{H$_{\rm o}$}
\def\q0{q$_{\rm o}$}
\def\lx{L$_{\rm X}$}
\def\l4{L$_{\rm X,44}$}
\def\nh{N(H{\sc i})}
\def\no{n$_{\rm e,0}$}
\def\p2c{P$_{\rm 20cm}$}
\def\sv{$\sigma_{\rm v}$}

\lefthead{HALL, ELLINGSON, AND GREEN}
\righthead{AGN HOST CLUSTER X-RAY EMISSION}
\slugcomment{Accepted to The Astronomical Journal, April 1997 issue}

\begin{document}

\title{X-RAY EMISSION FROM THE HOST CLUSTERS OF POWERFUL AGN}

\author{Patrick B. Hall}
\affil{Steward Observatory, University of Arizona, Tucson, AZ 85721 \\
E-Mail: pathall@as.arizona.edu}
\author{Erica Ellingson}
\affil{Center for Astrophysics and Space Astronomy, CB 391, University of Colorado, \\ Boulder, CO 80309}
\and
\author{Richard F. Green}
\affil{National Optical Astronomy Observatories\footnote{The National Optical 
Astronomy Observatories are operated by the Association of Universities for 
Research in Astronomy under Cooperative Agreement with the National Science 
Foundation.}, Tucson, AZ 85726-6732}

\begin{abstract}	\label{begin}

We report the detection of X-ray emission from the host cluster of the unusual
radio-quiet quasar \1821\ using the {\sl ROSAT} HRI, and the non-detection of 
X-ray emission from the host cluster of the radio-loud quasar 3C~206 (3$\sigma$\ 
upper limit of 1.63~10$^{44}$\ ergs~s$^{-1}$) using the {\sl EINSTEIN} HRI.
The host cluster of \1821\ is one of the most X-ray luminous clusters known,
with a rest-frame 0.1-2.4~keV luminosity of 
3.74$\pm$0.57~h$_{50}^{-2}$~10$^{45}$~ergs~s$^{-1}$, 
38\% of which is from a barely resolved cooling flow component.
The cluster emission complicates interpretation of previous
X-ray spectra of this field.
In particular, the observed Fe~K$\alpha$ emission can probably be attributed
entirely to the cluster and either the quasar is relatively X-ray quiet
for its optical luminosity or the cluster has a relatively low temperature
for its luminosity.

We combine these data with the recent detection of X-ray emission from the host
cluster of the `buried' radio-quiet quasar \9104\ (\cite{fc95}),
our previous upper limits for the host clusters of two z$\sim$0.7 
radio-loud quasars, and literature data on FR~II radio galaxies.
We compare this dataset to the predictions of three models for the 
presence and evolution of powerful AGN in clusters: the low-velocity-dispersion
model, the low-ICM-density model, and the cooling flow model.

Neither the low-ICM-density model nor the cooling flow model can explain all the
observations.
We suggest that strong interactions with gas-containing galaxies may be the only
mechanism needed to explain the presence and evolution of powerful AGN in
clusters, 
a scenario consistent with the far-IR and optical properties of the host
galaxies studied here, all of which show some evidence for past interactions.
However, the cooling flow model cannot be ruled out for at least some objects,
and it is likely that both processes are at work in creating and fueling
powerful AGN in clusters.  
Each scenario makes testable predictions for future X-ray and optical
observations which can test the relative importance of each process.

\end{abstract}


\section{Introduction}	\label{intro}

The association of quasars with galaxies at similar redshifts allows one to use
quasars as signposts for locating galaxies and galaxy clusters at high redshift.
Although radio-quiet quasars (RQQs) are very rarely found in clusters at 
z$\lesssim$0.7, a significant fraction of radio-loud quasars (RLQs) with 
0.4$<$z$<$0.7 are located in clusters of galaxies
(e.g., \cite{yg87} (YG87), \cite{eyg91} (EYG91), \cite{ey94}).  
The population of
quasars located in such rich clusters is seen to evolve 5--6 times faster than
their counterparts in poor environments (EYG91, \cite{ye93}): at z$\lesssim$0.4
RLQs are almost never found in rich clusters, at z$\sim$0.4--0.55 only faint
RLQs can be found in them, and at z$\gtrsim$0.55 both luminous and faint RLQs
are found there (cf. Figure 1 of \cite{ye93}).  This evolution can be 
extrapolated to include the very faint optical AGN activity seen in some radio 
galaxies in low-redshift rich clusters (e.g. \cite{dry90}).  
The environments of the population of Fanaroff-Riley class II (FR~II)
`classical double' powerful radio galaxies (PRGs)
evolve with redshift as well (\cite{pp88,ymp89,hl91,as93}).  
RLQs also have FR~II morphologies, and in the unification model (\cite{ant93})
PRGs and RLQs are the same class of objects seen at different orientations,
in which case their large-scale environments should be statistically identical.
One scenario which can explain these observations is that the physical
conditions in RLQ and PRG host cluster cores have undergone substantial changes
which have caused the high-z FR~II RLQs and PRGs in clusters to fade at optical 
wavelengths on a dynamical time scale and evolve into low-z optically 
faint FR~I radio galaxies.

In the context of this evolution the results of Tsai \& Buote (1996) are quite
interesting.  They conclude from hydronamical 
simulations that in an $\Omega$=1 CDM universe the formation rate of clusters 
from (unvirialized) background matter was large at z$\gtrsim$0.6, but dropped
off sharply at z$\sim$0.6 and remained constant and small at z$<$0.6, where new
clusters form primarily by mergers of preexisting, virialized smaller clusters.
This is very similar to the scenario suggested by Yee \& Ellingson (1993)
to explain the steep decline in the optical luminosities of RLQ population
in rich clusters at z$\lesssim$0.6.  
The RLQ luminosity function is consistent with a constant cluster birth rate
providing a continual supply of RLQ formation sites at z$>$0.6, in agreement
with the Tsai \& Buote simulations if clusters formed by mergers of 
preexisting subclusters are unfavorable sites for RLQ formation.
This agreement is intriguing but possibly coincidental, as other simulations
(\cite{rlt92,co94}) yield different cluster formation rates with redshift.

Many mechanisms have been put forth as explanations for the quasar
environment-evolution link:  galaxy-galaxy interactions and mergers
(\cite{hut84}), cooling flows (\cite{fc90}), and/or an intra-cluster medium
(ICM) of different density at high redshift (\cite{sp81}, \cite{bm88}).  
These different models have considerably different implications for X-ray 
observations of quasar host clusters, as discussed in Hall et~al. (1995;
hereafter Paper~I) and in \S\ref{disc}.  We are imaging quasars known to lie in
rich clusters with the {\sl ROSAT} High Resolution Imager (HRI; \cite{zom90})
to help discriminate among these scenarios for the evolution of powerful
AGN in clusters and the role played by the ICM in that evolution.
Paper~I presented upper limits for the first two quasars we studied.  
In this paper we add new data on one lower-redshift quasar,
archival data on another, and data from the literature on a third.
We then discuss the data in comparison to FR~II radio galaxy host 
clusters and optically and X-ray selected clusters.
We restrict our discussion primarily to PRGs and quasars with unambiguous FR~II
morphologies believed to be located at the centers of their host clusters.
Unless otherwise noted, we take \ho=50~\hf~\ksm, \qo=1/2, and $\Lambda$=0.

\section{Observations and Analysis}	\label{obsanal}

Table 1 
details the observations and results
discussed in this paper, Paper~I, and Crawford \& Fabian (1995).
\begin{table}
\dummytable\label{tbl-1}
\end{table}
The three objects newly added to our sample are discussed in detail in this
section.

\subsection{\9104} 	\label{9104}

\9104\ is a very IR-luminous `buried' radio-quiet quasar (\cite{hw93}; HW93) 
at the center of a rich, flattened cluster at z=0.442 (\cite{kle88}).
It is a radio source intermediate between FR~I and FR~II in
radio power and morphology (HW93).
HW93 show that \9104\ is powered by a hidden AGN based on
the detection of broad Mg~II $\lambda$2798 and strong wavelength-dependent
polarization.  
We classify \9104\ as a RQQ based on its k-corrected 5~GHz to 
(estimated) unobscured B luminosity ratio R$^*$ (\cite{sto92}).
Using data from HW93, we find R$^*$=0.89--4.27,
placing it at the high end of the range found for RQQs.  Also,
its position in the O[{\sc iii}]-P$_{5\rm GHz}$ plane (\cite{raw94}) clearly
shows it to be a RQQ, albeit an extreme one (similar to \1821), even after 
accounting for its steep radio spectral slope and correcting its large 
O[{\sc iii}] luminosity for non-nuclear emission (\cite{cv96}).
We estimated the richness \bgq\ of the host cluster CL~09104+4109 by using data
from Kleinmann et~al. (1988) to find N$_{\rm 0.5}$ (\cite{bah81}) and then
converting to \bgq\ using the empirical relation \bgq=34$\times$N$_{\rm 0.5}$ 
(\cite{hl91}).  We find \bgq=1210$_{-269}^{+316}$, equivalent to Abell richness
2.  The host galaxy is a cD possibly in the midst of 
cannibalizing several smaller galaxies (\cite{soi96}).
%

CL~09104+4109 has been detected by the \ROSAT\ HRI (\cite{fc95}; hereafter 
FC95; see also \cite{cv96}).  It is one of the most X-ray luminous clusters 
known and shows
evidence for a cooling flow in the form of an excess central emission component
above the best-fit King model.
An apparent deficit in the central X-ray emission is also observed,
possibly caused by H{\sc i} absorption in a cooling flow, 
or by displacement of the ICM by the radio jets 
or a mass outflow from the center of the host galaxy (\cite{cv96}).
FC95 calculate L$_{\rm X}$=2.9$\pm$0.25~10$^{45}$~ergs~s$^{-1}$
in the observed \rosat\ band, and kT=11.4$^{+\infty}_{-3.2}$~keV from a fit to
an ASCA spectrum of the object.  Using a kT=11.4~keV thermal brehmsstrahlung
spectrum redshifted to z=0.442, we calculate
a rest-frame 0.1--2.4~keV luminosity 3.03$\pm$0.26~10$^{45}$~ergs~s$^{-1}$.
Although the quasar and its cD host are at the center of the X-ray emission,
they may not lie at the optical center of the cluster (\cite{kle88}).


\subsection{3C~206}	\label{3C206}	

The first observation we present
is of the radio-loud quasar (RLQ) 3C~206 (z=0.1976).
3C~206 resides in a flattened cluster of Abell richness class 1 which has a
lower velocity dispersion than is typical for such clusters (\cite{ell89}).
The cluster (which we designate CL~3C~206) is very centrally concentrated,
with a best-fit optical core radius of 35~kpc (\cite{ell89}).
3C~206 is unusual in that it is the only radio-loud quasar at z$<$0.4 known to
reside in a cluster of Abell richness 0 or greater.
The host galaxy of 3C~206 is an elliptical in
the approximate optical center of the host cluster,
but the galaxy is $\gtrsim$1~mag fainter than expected for a first-rank
cluster galaxy (\cite{ell89,hut87}).
The host galaxy is slightly redder than a typical RLQ host galaxy but slightly
bluer than a normal elliptical (\cite{ell89}).

\subsubsection{Analysis of EINSTEIN HRI Observations of 3C~206}

Our `observation' of 3C~206 consists of a 61~ksec archival \ein\ High Resolution 
Imager (HRI) image.  No extended emission is obvious in the image.
Since our non-detection data analysis techniques were discussed in detail in
Paper~I, we give only an overview here.  Our modeling requires 
binned radial profiles for the object, the HRI Point Response Function (PRF),
and for PRF-convolved $\beta$=2/3 King model clusters of 
\rcore=125 and 250~kpc at the quasar redshift for several different cosmologies.
After background subtraction,
the PRF was normalized to the object counts in the innermost bin and subtracted,
leaving a radial profile consisting of any excess counts above the
profile expected for an object of the observed central intensity.
The object's radial profile, the fitted PRF and background, and the 
PRF subtracted (but not background subtracted) residual profile are plotted in 
Figure \ref{206fit}.
The residual is exaggerated in this log-log plot; note that the apparent excess
emission is of the same scale as the PRF and that the residual is negative
between 15-40\arcsec, where cluster emission should be most prominent.
Since the PRF fit is sufficient, we derive a \ctsec\ value for a 3$\sigma$
upper limit cluster as described in Paper I.
This upper limit on cluster \ctsec\ within 8\arcmin\ was corrected for deadtime
and vignetting through comparison with the \ein\ HRI source catalog (available
through the Einstein On-Line Service of the SAO).
We measured the counts of the quasar and the next brightest source in the 3C~206
field in exactly the same manner as the HRI source catalog.  
and found the archive 
count rates to still be a factor 1.112 higher than ours.  Although this deadtime
plus vignetting correction factor is somewhat higher than might be expected, 
we adopt it to be conservative.

To convert our limit from \ctsec\ to L$_{\rm X}$,
we first convert to the emitted flux in the \ein\ passband
corrected for Galactic absorption of log~N$_{\rm H}$=20.75 (\cite{elw89}).
We assume a Raymond \& Smith (1977) plasma spectrum with temperature from
\begin{equation}
\rm{\beta=\mu m_p \sigma_v^2/kT}	\label{A}
\end{equation}
where $\mu$ is the mean molecular weight of the cluster gas (0.63 for solar
abundance) and $\sigma_v$ is the cluster velocity dispersion.
Assuming $\beta$=2/3 gives kT=2.5$\pm$1.1~keV
using the observed $\sigma_v$=500$\pm$110 km~s$^{-1}$.
Using the \ein\ Users Manual (\cite{h84}), we find a conversion factor 
1.3~10$^{-13}$~ergs~cm$^{-2}$~count$^{-1}$ for this \nh\ and kT.

For intercomparison of our targets we convert to the rest-frame
\ROSAT\ passband (0.1-2.4~keV).
Conversions were calculated using redshifted thermal brehmsstrahlung
spectra and the effective areas as a function of energy for the \ein\ and
\rosat\ HRIs given in HH86 and David et~al. (1995).
%
Next we convert this rest-frame 0.1-2.4~keV band flux F to a luminosity
L$_{\rm X}$ using F=L$_{\rm X}$/4$\pi$D$_{\rm L}^2$, where D$_{\rm L}$
is the quasar's luminosity distance in the assumed cosmology:
\begin{equation}
\rm{D_L={{{2cz}\over{H_o(G+1)}}\left( {1 + {{z}\over{G+1}}}\right) }} \label{B}
\end{equation}
where G=$\sqrt{1+2zq_{\rm o}}$ (\cite{sg86}).
(Note that the equations for {\rm D$_{\rm L}$} given in Paper~I are incorrect.)
Finally, we correct the cluster luminosity for emission beyond r=8\arcmin.

The resulting upper limits, for different cosmologies and core radii,
are given in Table 2.
\begin{table}
\dummytable\label{tbl-2}
\end{table}
For kT=2.5~keV and r$_{\rm core}$=125~kpc
our CL~3C~206 3$\sigma$ upper limit is 1.63~10$^{44}$~ergs~s$^{-1}$.
Also given in Table 2 are our upper limits from Paper~I, now corrected to
the rest-frame 0.1-2.4~keV band.

\subsection{\1821} 	\label{1821}

\1821\ (z=0.297) is an IR-luminous X-ray selected radio-quiet quasar (RQQ)
residing in a giant elliptical galaxy
in a rich (Abell richness class $\sim$2) cluster at low redshift.
(\1821\ and \9104\ have the richest quasar host clusters known at any redshift.)
\1821\ has been detected and
studied in the radio despite being radio-quiet (\cite{lrh92,bl95,pap95,blu96}).
It has a core, a lobe, and two small jets (\cite{bl95,blu96}).
We calculate $\sigma_v$=1046$\pm$108 km~s$^{-1}$ for CL~1821+643 (in its rest
frame) using 26 members from Schneider et~al. (1992) and Le Brun, Bergeron 
\& Boiss\'e (1995).  
The host galaxy is bright, large, red, and featureless, but slightly
asymmetrical and offset from the nucleus by 1-2\arcsec\ (\cite{hn91b}).
All the galaxy's measured parameters are consistent with it being
a cD at the center of the cluster.

\subsubsection{Analysis of ROSAT HRI Observations of \1821}	\label{1821obs}

Our \rosat\ HRI observation of \1821\ is shown in Figure \ref{1821img},
binned into 1\arcsec\ pixels.  
Both the quasar and the nearby white dwarf central star of
the planetary nebula Kohoutek 1-16 (K1-16) are easily detected, along with
obvious extended emission from the quasar host cluster.  The X-ray emission
from K1-16 shows no signs of being resolved on our HRI image.  Isophote fitting 
from r=10\arcsec--70\arcsec\ on an adaptively smoothed (\cite{wbc95}) 
image showed that the cluster has an ellipticity of $\sim$0.1 at all radii.
The cluster isophotes' center is, within the errors, 
the same as the quasar position for all r$<$70\arcsec.
To decrease the FHWM and ellipticity of the PRF in our data, we subdivided the
image by exposure time, centroided, and restacked (cf. Morse 1994).
We also added a 1460~sec archival HRI image of the field at this stage.  
We then extracted the quasar and white dwarf radial profiles using annuli of 
1\arcsec\ width on the corrected, unbinned, unsmoothed HRI image, excluding
data within r=22\farcs5 of all objects detected by the standard processing,
and fitted the radial profile of emission surrounding the quasar.

\subsubsection{\1821\ Radial Profile}		\label{1821rp}

We initially assume a three-component radial profile:
a constant background, a \ROSAT\ HRI PRF, and a King surface brightness cluster
(${\rm S(r) \propto [1 + (r/r_{core})^2]^{-(3\beta-0.5)}}$)
convolved with the \ROSAT\ HRI PRF.
(Because of the complexity in fitting a non-analytic PRF-convolved King profile
to the data, we fit a simple King profile instead; simulations indicate $<$5\%
systematic error in this procedure, which we account for in our results.)
The parameters in our model are 
the background level, 
the normalization, core radius, and $\beta$ of the King profile,
the PRF normalization,
the widths $\sigma_1$ and $\sigma_2$ of the two gaussians that comprise the PRF
(see Paper~I and David et~al. (1995); hereafter D95),
and the relative normalizations of the PRF gaussians.
We kept the normalization and scale length of the exponential PRF component 
fixed at the standard values in all models.

We used {\sc nfit1d} in {\sc IRAF}\footnote{The 
Image Reduction and Analysis Facility 
is distributed by National Optical Astronomy Observatories,
operated by the Association of Universities for Research in Astronomy, Inc.,
under contract to the National Science Foundation.}
to fit our model to the observed radial profile.  
We used $\beta$=2/3 and the standard normalization of the two PRF gaussians;
all other parameters were allowed to vary.
The solution (plotted in Figure 3a)
was good (reduced $\chi^2$=0.8612)
but it gave a value of $\sigma_2$=4\farcs58$\pm$0\farcs05, noticeably above the
maximum of 4\farcs1 measured for long PRF characterization observations
(\cite{hri95}).
We attempted to produce an adequate fit more in line with the known PRF 
properties by allowing the all the parameters to vary, but no fit gave
a smaller $\sigma_2$.

To see if our broad PRF result was robust, we fitted the PRF of the white dwarf
using data within r=23\arcsec.  
The best fit (reduced $\chi^2$=0.698) had $\sigma_1$=2\farcs03$\pm$0\farcs18 
and $\sigma_2$=4\farcs08$\pm$0\farcs20, consistent with the standard values
and the range of observed values (D95).
So, either the PRF in our image is best given 
by the fit to the quasar, not the white dwarf, and is slightly broader than 
measured in the PRF characterization observations, or the PRF is best given by 
the fit to the white dwarf and there is a barely resolved component to the 
cluster emission which broadens the fitted quasar radial profile.
Such a component would most likely be due to a cooling flow.

To test the hypothesis that a single King profile is inadequate to
describe the cluster emission,
we fitted a simple gaussian along with a point source and a $\beta$=2/3
King profile.  The best fit (plotted in Figure 3b) was a considerable improvement 
(F-test probability $<$0.5\% of occuring by chance) and had 
$\sigma_2$=2\farcs34$\pm$0\farcs04 and $\sigma_2$=3\farcs91$\pm$0\farcs20, 
smaller than in the fit without the extra gaussian and consistent with the WD
fit and the range observed in PRF characterization observations.  However,
the amplitude of the gaussian component is constrained to only $\pm$32\%.

Thus the total cluster emission is better described by a King profile plus a
barely resolved component than a King model alone, but the amplitude of the 
barely resolved component is considerably uncertain.  
As a check, we fitted a King model and gaussian to the original, uncorrected
image, and found that the parameters for both components were identical within
the errors to the values determined from the corrected image.
The King component of the cluster's total flux is 3523$\pm$498 counts.
The best-fit additional gaussian component has 2139$\pm$713 counts,
for a total of 5662$\pm$870 cluster counts, integrated to infinity.

\subsubsection{CL~1821+643 Physical Parameters}		\label{1821pp}

Several steps must be taken to convert from \ctsec\ to L$_{\rm X}$.
We take Galactic log~\nh=20.58 ({\cite{ls95}) and assume a Raymond-Smith 
spectrum with observed kT=5~keV, the highest value tabulated in D95;
a higher value would increase the estimated \lx\ only a little,
since \rosat\ has little effective area above 2~keV.
We find the energy-to-counts conversion factor as described in Paper~I and
divide this factor (0.223) into our \ctsec\ limit to obtain the energy flux 
in units of 10$^{-11}$ \cgsflux.
We then convert to L$_{\rm X}$ in the rest-frame 0.1-2.4~keV band.  
We measure a rest-frame 0.1-2.4~keV luminosity of
3.74$\pm$0.57 h$_{50}^{-2}$~10$^{45}$~ergs~s$^{-1}$ for CL~1821+643
with 2.33$\pm$0.33 and 1.41$\pm$0.47 h$_{50}^{-2}$~10$^{45}$~ergs~s$^{-1}$
from the King model and cooling flow components respectively.
The values of \lx\ for different cosmologies are tabulated in Table 2.
This luminous cluster complicates the interpretation of previous X-ray
observations of this field (Appendix \ref{yaq}).

The detection of ICM X-ray emission allows us to calculate the central
electron number density \no\ of the cluster
if the emission follows a King model, and to put a lower limit on it 
in the case of a cooling flow.
For $\beta$=2/3, equation (3) of Henry \& Henriksen (1986; HH86) becomes:
\begin{equation} 
\rm{I(0; E_1 , E_2)=1.91 \times 10^{-3} n_{e,0}^2 r_{c} \sqrt{kT}~[\gamma(0.7,E_1/kT) - \gamma(0.7,E_2/kT)]~~ergs~s^{-1}~cm^{-2}~sr^{-1}}	\label{D}
\end{equation}
where 
r$_{\rm c}$ is the cluster core radius in kpc,
\no\ is the central electron number density of the cluster in cm$^{-3}$,
kT is the cluster temperature in keV,
and $\gamma$(a,z) is the incomplete gamma function 
$\int_z^\infty$~x$^{a-1}$~e$^{-x}$~dx.
(The order of the gamma function terms is incorrectly reversed in HH86).
I(0;E$_{\rm 1}$,E$_{\rm 2}$) is the cluster's central X-ray surface brightness
(at the cluster) in the band E$_{\rm 1}$ to E$_{\rm 2}$,
and E$_{\rm 1}$ and E$_{\rm 2}$ are the energies (in keV) corresponding to the
lower and upper limits, respectively, of the instrumental passband
{\it at the object's redshift}.
For \rosat, E$_{\rm 1}$=0.1(1+z)~keV and E$_{\rm 2}$=2.4(1+z)~keV.
I(0;E$_{\rm 1}$,E$_{\rm 2}$) can be related to I$_{\rm obs}$, the observed
central surface brightness in the (E$_{\rm 1}$,E$_{\rm 2}$) band 
in \cgsflux~arcsec$^{-2}$ as follows (see also \cite{bw93}).
For $\beta$=2/3, the total cluster X-ray luminosity in the
(E$_{\rm 1}$,E$_{\rm 2}$) band is easily found by 
integrating the surface brightness either at the source or at the observer.  
Equating the two, we have:
\begin{equation}
{\rm L_X(E_1 , E_2) = I(0; E_1 , E_2)~4\pi~2\pi r_{core}^2 = I_{obs}~4\pi d_L^2~2\pi\theta_c^2}		\label{E}
\end{equation}
where r$_{\rm core}$ and d$_{\rm L}$ are in cm
and $\theta_c$ is the angular size corresponding to \rcore\ at the object's
redshift.  This yields
\begin{equation}
{\rm I(0; E_1 , E_2) = I_{obs}~d_L^2~\theta_c^2/r_{core}^2}	\label{F}
\end{equation}
Also, if the cluster T is unknown but $\sigma_v$ is, we can use Eq.~\ref{A}
for $\beta$=2/3 to replace $\sqrt{\rm kT}$ in Eq.~\ref{D}:
\begin{equation} 
\rm{I(0; E_1 , E_2)=5.86 \times 10^{-6} n_{e,0}^2 r_{c} \sigma_v~[\gamma(0.7,E_1/kT) - \gamma(0.7,E_2/kT)]~~ergs~s^{-1}~cm^{-2}~sr^{-1}}	\label{G}
\end{equation}
where $\sigma_v$ is in km~s$^{-1}$.
For \1821, we find 
\no=0.015$\pm$0.002 h$_{50}^{1/2}$ cm$^{-3}$ for the King model component
and a lower limit of \no=0.081$\pm$0.022 h$_{50}^{1/2}$ cm$^{-3}$ for the 
cooling flow component using the central surface brightness of our gaussian fit.
(Strictly speaking a deprojection analysis is required to calculate \no\ in the
case of a cooling flow, and the densities will still be underestimated because
the central regions are unresolved,
but our estimates should be accurate lower limits.)

With an estimate for \no, we can estimate \tcool, the cooling time for gas in
the center of the cluster, using equation (5.23) of Sarazin (1988):
\begin{equation}
{\rm t_{cool}=2.89~10^7~n_{e,0}^{-1}~\sqrt{T}~years}	\label{H}
\end{equation}
where T (estimated from \sv\ using Eq.~\ref{A} if necessary) is in keV,
\no\ is in cm$^{-3}$, and we have used the relation 
n$_{\rm p}$=0.82\no\ for completely ionized H-He gas.  We find
\tcool$<$6.4$\pm$1.2 h$_{50}^{-1/2}$~Gyr (since \no$\propto$h$_{50}^{1/2}$),
which is less than the age of the universe at z=0.297, 
8.8~h$_{50}^{-1}$~Gyr (10.1~h$_{50}^{-1}$~Gyr for \qo=0),
for all reasonable \ho.
Thus CL~1821+643 meets the standard criteria for
the presence of a central cooling flow.
%
The mass cooling rate \mcool\ can be found from 
\begin{equation}
{\rm L_{cool} = 2.4\ 10^{43}\ T_{\rm keV}\ \dot{M}_{\rm cool,100}\ ergs\ s^{-1}}   \label{I}
\end{equation}
where L$_{\rm cool}$ is the total luminosity of the cooling gas,
T$_{\rm keV}$ its initial temperature, 
and \.{M}$_{\rm cool,100}$ the mass deposition rate
in 100 h$_{50}^{-2}$ M$_{\sun}$ yr$^{-1}$ (\cite{fab86}).
We find \mcool=1120$\pm$440 h$_{50}^{-2}$ M$_{\sun}$~yr$^{-1}$ for \1821.  This
is likely a lower limit since we have not used the bolometric L$_{\rm cool}$.

\section{Discussion}	\label{disc}

\subsection{Physical Parameters of Quasar Host Clusters}	\label{phys}

Only five quasar host clusters have observations deep enough to put 
interesting limits on their X-ray emission.
The two detections and three upper limits are listed in Table 2.
We note that the three upper limits are for the host clusters of RLQs 
with P$_{\rm 20cm}$$>$10$^{26}$ W/Hz and the detections are of 
the host clusters of two RQQs with P$_{\rm 20cm}$$\sim$10$^{25}$ W/Hz,
two of the richest quasar host clusters known at any redshift, 
which are two of the most X-ray luminous clusters known.

These two luminous RQQ host clusters have dense ICM 
(cf. Table 1, \S\ref{1821pp}).
For CL~09104+4109,
using the best-fit FC95 King model 
\rcore=30\arcsec, kT=11.4$\pm$3.2~keV, and an extrapolated central 
surface brightness of 0.3--4.0~counts~arcmin$^{-2}$~s$^{-1}$ 
(see Fig. 3 of FC95), we obtain from Eq.~\ref{D} a lower limit for
\no\ (electron density at the cluster center)
in the range 0.027--0.097 h$_{50}^{1/2}$ cm$^{-3}$.
This is consistent with the value of $\sim$0.038 h$_{50}^{1/2}$ cm$^{-3}$ at
r$<$50~kpc obtained from the deprojection analysis of Crawford \& Vanderreist
(1996), but we note once again that these values are upper limits for the true
central densities, because of the cooling flows.
Cooling flow gas should have roughly r$^{-1}$ density and pressure profiles
(\cite{fab94}, p. 299), and the X-ray images do not resolve the innermost
regions where the density should be highest.
In addition, the apparent central X-ray deficit in
\9104\ may alter conditions in the cluster center.
For comparison, Abell clusters typically have
\no=0.001-0.010 h$_{50}^{1/2}$~cm$^{-3}$ (\cite{jf84}), 
and the host clusters of the radio galaxies Cygnus~A and 3C~295 have 
\no=0.057$\pm$0.016 
and $>$0.026$_{-0.009}^{+0.018}$ h$_{50}^{1/2}$ cm$^{-3}$ respectively 
(\cite{cph94}, HH86).  

We can also constrain \no, \tcool, and \mcool\ for the three RLQ host clusters.
Using our upper limit surface brightnesses for \rcore=125~kpc and assuming 
kT=5~keV (kT=2.5~keV for 3C~206), we obtain the limits shown in Table 1.
The limiting central surface brightnesses and density limits are
lower and the cooling times longer for \rcore$>$125~kpc and/or \ho$>$50,
but can be shorter if the gas is abnormally cool or centrally concentrated
(e.g. in galaxy size halos or clusters with \rcore$<$125~kpc).
For 3C~206, the cooling time at the center of a putative cluster right at our
3$\sigma$\ upper limit
is less than the age of the universe at z=0.1976 for plausible cosmologies.
For 3C~263 and PKS~2352-342, if \qo=0.5 the age of the universe at their 
redshifts is several Gyr shorter than the cooling times of their host clusters
and thus no cooling flows are possible, but cooling flows are possible if \qo=0.
If we assume \qo=0 and make a maximal assumption of clusters just at 
our 3$\sigma$\ upper limits with 50\% of their emission from cooling flow 
components, we obtain the limits on \mcool\ shown in Table 1.
We discuss the implications of these values later, in \S\ref{cf}.

Another interesting constraint on some of the clusters' line-of-sight
properties can be made.  
\1821\ has significant flux below 912~\AA\ (\cite{kol91,lee93,kri96}); thus,
the cooling flow in CL~1821+643 does not produce a Lyman limit and must have
intrinsic \nh$\leq$10$^{17}$~cm$^{-2}$ along the line of sight.  This is also
the case for CL~3C~263 (\cite{cra91}) but not for CL~09104+4109, which has
intrinsic \nh=2.5$_{-1.1}^{+1.8}$ 10$^{21}$~cm$^{-2}$ (FC95).  
This latter value is typical for nearby cooling flows 
(\cite{aea95}, but cf. Laor 1996);
thus, the cooling flow in CL~1821+643 and any putative cooling flow in CL~3C~263
have unusually low intrinsic \nh\ along our line of sight.
Either the cooling flows have low overall \nh, or,
more likely, the ionizing radiation from the two quasars is 
confined to a cone (including our line of sight) within which cooling 
gas is reionized, as suggested by Bremer, Fabian \& Crawford (1996).

\subsection{Comparison of Observations and Models}	\label{compare}

In this section we compare our observations 
to three models which have been proposed to explain the evolution of AGN cluster
environments.  We introduce each model, discuss data on key predictions, point
out problems, and discuss implications and possible solutions to the problems.

\subsubsection{The Cooling Flow Model}		\label{cf}

The cooling flow model (\cite{fab86,fc90}) is not a model for quasar formation
in cooling flows, but rather a model for how dense cooling flows can fuel
AGN located within them in a self-sustaining manner.  However, if it is to
explain the evolution of RLQs and PRGs in clusters at z$<$0.6, 
such objects must preferentially be found in cooling flow clusters for some reason,
perhaps because radio galaxies in clusters have higher radio luminosity
due to radio lobe confinement (\cite{ba96}).
This model is supported by the work of Bremer et~al. (1992, and references 
therein), who find that extended line-emitting gas around z$\lesssim$1 RLQs
is so common that it must be long-lived and therefore confined.  
If a hot ICM confines the gas, the required pressure is such 
that the ICM should have cooling flows of 100-1000~\myr.  
Fabian \& Crawford (1990) outline a model where luminous quasars at z$>$1 
are surrounded by dense cooling flows in subclusters.  
They show that an AGN of luminosity L in 
dense (P=nT$\sim$10$^8$ K/cm$^3$) gas at the virial temperature of the central
cluster galaxy (T$\sim$10$^7$~K)
will Compton-cool the gas within a radius R$\propto$L$^{1/2}$ for a
mass accretion rate (proportional to this volume) of \mdot$\propto$L$^{3/2}$.
Since L$\propto$\.{M}c$^2$, this Compton-cooled
accretion flow will grow by positive feedback until L=L$_{\rm Edd}$, 
as long as Compton cooling dominates brehmsstrahlung for L$<$L$_{\rm Edd}$,
which occurs only in high-P environments.  
AGN so powered will last until a major subcluster merger
or until the supply of dense cooling gas is exhausted.  
Assuming the most luminous objects form in the densest regions, the observed
optical fading of the RLQ/PRG population in clusters below z$\sim$0.6 can be 
explained by major subcluster mergers (which occur earlier in the richest 
environments) disrupting the Compton-cooling mechanism, leading to a precipitous
drop in the quasar luminosity.  
Thus this model is also intriguingly consistent
with the simulations of Tsai \& Buote (1996) discussed in \S\ref{intro}.

The cooling flow model can be tested in detail for 3C~263.
Crawford et~al. (1991), hereafter C91, predict \mcool=100-1000~\myr\ for
CL~3C~263 from extended emission-line gas observations.
We predict at most \mcool$<$202~\myr\ for CL~3C~263, and then only if \qo=0.
For \qo=0.5, to have \tcool\ less than the age of the universe at its 
redshift and \mcool=100~\myr, CL~3C~263 must have 
either \rcore$<$62~kpc and cooling flow \l4=1.2, or T$<$1.3~keV and cooling 
flow \l4=0.3, where \l4\ is X-ray luminosity in units of 10$^{44}$~erg~s$^{-1}$.
In addition, the minimum energy pressure 100~kpc 
from the quasar given by C91 can be produced by the ICM only if there is
a cluster with kT=5~keV and \rcore=125~kpc right at our upper limit luminosity.
However, matching the pressure at $<$100~kpc inferred by C91 from
any of their models based on observed [O{\sc iii}]/[O{\sc ii}] line ratios and
various assumptions for the quasar's ionizing spectrum requires an additional 
compact cooling flow component or a cluster with \rcore$\lesssim$100~kpc.
Thus for CL~3C~263 to match the predictions of the cooling flow model as given 
in C91, it cannot be much fainter than our upper limit and must have 
\rcore$\lesssim$60--100~kpc and/or an unusually low kT for its \lx.
Crawford \& Fabian (1989) and Fabian (1992) point out, however, that since
collapsed structures at high z have shallower potentials, the gas in them will
have a lower kT, and since they are denser, ``more compact objects 
than present-day clusters may be appropriate sites for remote cooling flows.''

The cooling flows in our two RQQ host clusters may very well have central ($<$1~kpc) pressures $\sim$10$^8$~K/cm$^3$, and thus be explained by the cooling flow
model's Compton feedback mechanism, again with the caveat for \9104\ that 
the apparent central X-ray deficit may indicate unusual conditions.
But if pressures do reach 10$^8$ K/cm$^3$ at $<$1~kpc in massive
(\mdot$\gtrsim$500~\myr) cooling flows at low redshift (z$\lesssim$0.4), 
this model must explain why surveys of powerful AGN at such redshifts 
very rarely find them in massive cooling flow clusters.
Also, if our three RLQ host clusters have cooling flows at all, 
they would have \mdot$\lesssim$200~\myr.
Central pressures $\sim$10$^7$~K/cm$^3$ have been estimated for cooling flows
of this strength at low redshift (\cite{heck89}).
Thus the central pressures in these RLQ host clusters are not likely to be
consistent with the cooling flow model 
because their central pressures are an order of magnitude lower 
than required for the Compton-cooling feedback mechanism of Fabian \& Crawford 
(1990) to successfully power the AGN.
{\it This evidence suggests that the cooling flow model cannot be the sole
explanation for the evolution of powerful AGN in clusters.}
However, current data is insufficient to completely rule the model out as the
sole explanation, since our three RLQ host clusters which seem to lack dense
cooling flows {\it might} be powered in the manner predicted by this model
{\it if} 1) cooling flows are in more compact and cooler clusters at high redshift
or 2) there is something unusual about these objects which has caused the 
density of hot gas in their innermost regions to increase by at least an order
of magnitude above the density predicted by X-ray data (cf. \S\ref{spec}).

As for powerful FR~II radio galaxies (PRGs), at low redshift they are extremely
rarely found in clusters (which preferentially host FR~I radio galaxies), 
and most of those that are in clusters are not at the cluster centers 
(\cite{lo95,wd96}), which may in itself argue against the cooling flow model.
The only two clusters with central FR~II galaxies in which cooling flows could 
have been definitively detected with observations of the S/N and spatial 
resolution made to date are CL~3C~295 and CL~CygA, both of which have cooling 
flows of $\sim$200~\myr\ (see references in legend to Fig. \ref{LvB}).
The next best candidate for a FR~II radio galaxy at the center of 
a cooling flow is B3~1333+412 in A1763 at z=0.189 (\cite{vb82}).
For CL~3C~295, existing observations do not rule out high central pressures.
For CL~CygA, Reynolds \& Fabian (1996) find P=2.5~10$^6$~K/cm$^3$ at 15 kpc.
This extrapolates to $\sim$4~10$^7$~K/cm$^3$ at r=1~kpc assuming 
P$\propto$r$^{-1}$ (\cite{fab94}).  
So it is possible these two PRGs have central densities and pressures sufficient
to support the Compton feedback quasar fueling mechanism, but it is also
possible that some other process is needed to create the required high densities.

The finding of Wan \& Daly (1996) that FR~II host clusters at z$\leq$0.6 are
typically X-ray underluminous (i.e. cooler or less dense than average clusters)
may also be a problem for this model (cf. Fig. \ref{LvB}).
The central densities they give 
clusters translate into cooling times longer than the age of the universe for
all clusters in their sample except CL~CygA.
But Wellman, Daly \& Wan (1996a, 1996b), using radio bridge 
parameters of a sample of FR~II radio galaxies at z=0.5--1.8,
derive somewhat larger surrounding densities and temperatures,
consistent with present day ICM, and
cooling times short enough to form cooling flows in some cases (cf. \S\ref{icm}).

Thus the cooling flow model for the evolution of FR~II AGN environments is
unlikely to be the {\it sole} explanation for this evolution.  \1821\ and \9104\
are easily explained by this model, but all other host clusters we have
discussed may harbor cooling flows as dense as required by the model only if
1) cooling flows are found in cooler and denser clusters at z$\gtrsim$0.4,
or 2) some other mechanism has boosted their central densities 
high enough to engage the fueling mechanism proposed in the model.
(However, Cygnus~A and 3C~206 need not follow this model for it to explain the 
evolution of FR~II AGN environments at z$\gtrsim$0.4.)
One possible mechanism for increasing central densities is strong interactions
or mergers, discussed in \S\ref{spec}.

\subsubsection{The Low ICM Density Model}	\label{icm}

The low-ICM-density model (\cite{sp81}, EYG91, {\cite{ye93}) predicts that
quasars are preferentially located in host clusters with low-density ICM
($\lesssim$10$^{-4}$ cm$^{-3}$) where ram pressure stripping 
is inefficient and gas remains in galaxies as a possible AGN fuel source.
This model is consistent with 
findings that radio sources at z$\sim$0.5 have radio morphologies 
uncorrelated with the richnesses of their environs (\cite{rse95,hut96}), 
implying that at z$\sim$0.5 the ICM/IGM in optically rich environments is not
consistently denser than in poor ones.  
Similarly, Wan \& Daly (1996) 
find that at z$<$0.35 clusters with FR~II sources tend to be less 
X-ray luminous (less dense and/or cooler) than those without.
FR~II host clusters at z$\sim$0.5 are consistent with being underluminous 
as well, based on comparison of inferred radio bridge pressures to those in
the z$<$0.35 sample.
However, this model is inconsistent with
Wellman, Daly \& Wan (1996a, 1996b), who find from radio bridge studies
that at z=0.5--1.8 large FR~II 3C radio galaxies may be surrounded by gas with 
densities and temperatures similar to nearby clusters.  
These different results may be explained by the different radio powers
of the objects in each sample (cf. \cite{ba96}).
Also, some high-z radio galaxies are observed to have large 
rotation measures which at low z are caused only by dense ICM (\cite{car97}).

This model predicts that powerful AGN host clusters are
X-ray underluminous for their optical richnesses. 
In Figure \ref{LvB} we plot cluster richness B$_{\rm gc}$ vs. rest-frame
soft X-ray luminosity L$_{\rm X}$(0.1-2.4~keV) to look for this trend.
B$_{\rm gc}$ 
is a linear measure of richness.
Quasar host clusters are plotted as filled squares,
with upper limits assuming r$_{\rm core}$=125~kpc.
Open squares are a \={z}$\sim$0.3 subsample of X-ray 
selected EMSS clusters studied by the CNOC group (\cite{car96}),
radio galaxies are filled triangles, and other symbols are objects from the
literature 
(see figure legend for references).
The dotted line is the best-fit relation to the CNOC/EMSS data.
Compared to both the CNOC sample and other clusters from the literature,
\1821\ and \9104\ are X-ray overluminous for their optical richnesses
and 3C~206, 3C~263, and \2352\ are either normal or underluminous,
consistent with this model.
Note, however, that several z$>$0.5 optically selected clusters (half-filled
squares; see figure legend for references) lie at the low-\lx\ end of the 
literature range; thus, our two high-z RLQ host clusters might be normal or even
overluminous compared to high-z optically selected clusters of similar richness.

Clusters with central FR~II radio galaxies (filled triangles)
are either normal or underluminous for their richnesses.
The \lx\ for 3C~382 is probably contaminated by the central engine,
as an archival WFPC2 image reveals a bright, unresolved source in the
center of the host galaxy.  
Thus the only overluminous PRG host clusters are those of 3C~295 and Cygnus~A,
but these objects do present immediate problems for this model.
Perley \& Taylor (1991) argue convincingly that 3C~295 is $<$10 Myr old, based
on its observed size and the ram pressure needed to confine the hot spots.
Similar calculations for Cygnus~A 
yield an age of $\sim$30-40~Myr (\cite{car91,bla96}).
Multiple-generation AGN models predict characteristic lifetimes of 
$\sim$100~Myr and single-generation models even longer ones (\cite{cp88}),
so these truly are young AGN.
If the low-ICM-density model is correct and radio sources should not form in
dense clusters, these
clusters must have grown dense only after the radio sources formed.
However, the shortest timescale on which the ICM might 
significantly evolve is the cluster-core sound crossing time, $\sim$100 Myr, so
it appears some strong radio sources have formed while immersed in dense ICM.

The existence of these two high ICM density RQQ host clusters 
is a problem for the low ICM density model if it is to be a
universal explanation for the presence and evolution of quasars in clusters.  
But as RQQs in clusters are rare, they may very well have different 
evolutionary histories than RLQs in clusters.  One possibility is that the RQQs
formed as RLQs when the clusters were less dense and have been
continuously active ever since.  
Their evolution into RQQs sometime after formation
might have been due to spin-down of the black hole
(\cite{bec94}) or the increasing density of their environments
interfering with jet production (\cite{ree82}).  

In this scenario,
even if we assume the ICM density doubles on a dynamical timescale, a rate 
ten times faster than simulations predict (\cite{evr90}) 
but probably still consistent with data on high-z optically selected 
clusters (\cite{cas94}), \1821\ and \9104\ must be quite old ($\sim$10$^9$~yr)
to have formed in clusters with even moderately low ICM densities
(\no$\lesssim$10$^{-3}$~cm$^{-3}$).
If these two RLQ host clusters are typical of z$\sim$0.7 RLQ host clusters,
then with this assumed rapid evolution of the ICM density it is possible that
these clusters could evolve to be as luminous as the two RQQ host clusters by
z$\sim$0.30--0.44, consistent with the suggestion that these RQQs were once RLQs.
Another constraint comes from the mass of the central black hole in \1821,
which is estimated at M$_{\rm BH}$=3~10$^9$ M$_{\sun}$ (\cite{kol93}).
If \1821\ has accreted continuously at the Eddington limit with a 10\%
efficiency for conversion of accreted matter into energy, it would have
reached this M$_{\rm BH}$ after 0.9~Gyr, just consistent with the age necessary 
for formation in a low ICM density cluster.
If the efficiency were any less, the black hole would have reached its estimated
mass more quickly and the quasar would have to be younger.  If the accretion
rate were any less, the quasar would not likely be as luminous as it is.

Thus while these two RQQs could be very old continuously active quasars which
formed as RLQs in moderately low ICM density clusters,
the required rate of ICM density increase is extremely large,
the different timescales involved agree for only a small range of parameters,
and the requirement for continuous fueling of these rather luminous quasars at
the Eddington rate for $\sim$1~Gyr is a difficult one to fulfill without
invoking interactions which allow gas to flow into the center of the host 
galaxies.
This scenario does make the testable prediction that if any remnant radio lobes
exist around these objects, they should be very old.

The major drawback of this scenario is that it is not applicable to the two PRGs
(Cygnus~A and 3C~295), since those AGN are very young.  One explanation which
might be applicable to all four objects in high ICM density clusters is that they
were created recently when their host galaxies underwent strong interactions.
We defer discussion of this possibility to \S\ref{spec}.
In any case, {\it the existence of these four objects in high ICM density 
clusters is sufficient to rule out the low-ICM-density model as the sole
explanation for the presence of powerful AGN in clusters,}
even though current data do not rule out 
low density ICM being present in {\it most} powerful AGN host clusters.

\subsubsection{The Low-$\sigma_v$ Interaction/Merger Model}	\label{sigv}

The low-\sv\ interaction/merger model (\cite{hut84}, EYG91) predicts that
quasars will be preferentially found in unvirialized, low velocity dispersion
($\sigma_v$) clusters where the low-$\Delta$v encounters needed for
strong interactions and mergers are relatively common (\cite{af80}).
Ellingson, Green \& Yee (1991) showed that the composite 
$\sigma_v$ of quasar host clusters is
lower than for comparably rich Abell clusters, consistent with this model.

To test this model, in Figure \ref{DvB} 
we plot cluster richness \bgq\ vs. cluster velocity dispersion $\sigma_v$.
Compared to the CNOC and literature data, 3C~206 has a slightly low \sv\ for its
richness while \1821\ is normal.  
The $\sigma_v$ and \bgq\ of 3C~206 are representative of the ensemble quasar
host cluster of Ellingson, Green \& Yee (1991).
The host clusters of the PRGs Cygnus~A and 3C~295 are outliers.  
Smail et~al. (1997) give 1670~\kms\ (21 objects) for CL~3C~295,
and $\pm$250~\kms\ uncertainty (Smail, personal communication).
The redshift histogram of CL~3C~295
shows no evidence for subclustering or field contamination (\cite{dg92}).  
The $\sigma_v$ of CL~CygA, based on only five galaxies,
is almost certainly an overestimate (\cite{ss82}).

Thus the few available measurements of quasar and FR~II radio galaxy host
cluster velocity dispersions are not particularly supportive of this model,
although some measurements may suffer from field contamination (\cite{sma97}).
Of the two RQQs in our sample, no \sv\ measurement exists for CL~09104+4109,
and CL~1821+643 has a normal or high $\sigma_v$ for its richness.
Since the cluster \sv\ evolves on the dynamical timescale during formation,
we can make the same arguments about the age of \1821\ 
as we did in the previous discussion for the low-ICM-density model, namely that
these RQQs could be old AGN which formed as RLQs when the cluster had a lower \sv.  
But for 
CL~3C~295 and CL~CygA, even if we assume $\sigma_v$$\sim$850~km~s$^{-1}$ 
for consistency with X-ray data (\cite{hh86,car91}) their velocity dispersions
would still be normal or high for their richnesses, and these AGN are too
young to have formed when their clusters had lower \sv.
As we discuss in the next section, however, there is another possible
explanation for these exceptions to the low-\sv\ model which may preserve
the model's viability.

\subsection{Which Model(s) Are Correct?} \label{spec} 

None of the three simple models purporting to explain the presence and evolution
of powerful 
AGN in cluster centers seem able to explain all the observations at first glance.
The low ICM density model cannot account for AGN in clusters with dense ICM 
(3C~295 \& Cygnus~A, and \1821\ \& \9104\ unless they are very old; c.f. \S\ref{icm}),
but is consistent with our nondetection of ICM emission from RLQ host clusters.
Some other recent radio and X-ray work supports this model 
(\cite{rse95,hut96,wd96}), but some does not 
(Wellman, Daly \& Wan 1996a, 1996b; \cite{cf96b}).
The cooling flow model requires very strong cooling flows, and thus has
difficulty accounting for host clusters without luminous X-ray emission
or with relatively weak cooling flows, 
but can explain the X-ray luminous host clusters of \1821\ and \9104.
And the low-$\sigma_v$ model has difficulty explaining 3C~295, \1821,
and Cygnus~A, whose host clusters have apparently high $\sigma_v$'s,
although it is supported by the scarce data on RLQ host clusters (\cite{eyg91}).

The evidence suggests that neither the cooling flow nor the low-ICM-density
models can be the sole explanation for the presence and evolution of powerful 
AGN in clusters.
Strictly speaking, the low-\sv\ model cannot be the sole explanation either,
since some powerful AGN reside in high-$\sigma_v$ clusters.
However, even in high-$\sigma_v$ clusters, the low-$\Delta$v interactions or
mergers required by the low-\sv\ model can still occur, albeit rarely.
{\it We suggest the possibility that AGN are produced 
in clusters solely by a strong interaction of their host galaxy with another
galaxy in the cluster.}  (We define a strong interaction as an interaction 
and/or merger which results in considerable gas flow into the central regions
of the post-interaction AGN host galaxy.)  
This would naturally favor host clusters with low \sv\ (and possibly low ICM
density), since strong interactions with gas-containing galaxies are more common
in such clusters, but again, such interactions can occur in any cluster
(\cite{dm94}) as well as in the field where most quasars exist.
If this strong interaction scenario is correct,
the distribution of AGN host cluster velocity dispersions should be biased to 
low values, but need not consist exclusively of low-\sv\ clusters.
A large sample of AGN host cluster \sv's will be needed to test that prediction.
A more easily testable prediction is that all host galaxies of quasars in 
clusters should show evidence of interaction with another galaxy.

It is also possible that strong interactions are not the sole formation process
for AGN in clusters, and that the cooling flow 
model operates in some cases.
Strong interactions may also allow the cooling flow model to operate in clusters
it would not otherwise be able to, by providing a mechanism for increasing the
ICM densities at the center of the host galaxies sufficiently high to switch on
the Compton-feedback fueling mechanism.
X-ray observations of additional powerful AGN in clusters will determine how
prevalent the cooling flow model can be, and how necessary an additional
mechanism for increasing the central densities is.
Observations of the host galaxies of such quasars will determine how often
mergers might provide that mechanism.

We now consider whether there is evidence for or against this strong 
interaction scenario in the far-IR and optical properties of the host galaxies
of the AGN we have discussed (cf. Table 1).  

\paragraph{Far-IR Properties:} \label{farir}
{\bf RQQs:} Both \9104\ and \1821\ are luminous far-IR sources (Table 1),
with a 60$\mu$m to optical luminosity ratio at least 2.5 times greater than
any of our three RLQs.
This excess far-IR emission above what is expected for quasars of 
their luminosity can plausibly be attributed to an excess of gas and
dust in the RQQ host galaxies resulting from a recent strong interaction.
The excess IR luminosity is too strong to be attributed to gas and dust in the
cooling flow (\cite{bmo90}).
{\bf RLQs:} 
The far-IR luminosity of 3C~206 is almost two orders of magnitude lower than
the two RQQs.
Both 3C~263 and \2352\ cannot be ruled out as being IR-luminous, although at
most they would still be an order of magnitude less luminous than the two RQQs.
Thus any interactions in which these RLQs were involved must have been much less
gas-rich than those in which the RQQs were involved.
{\bf PRGs:} 3C~295 has log~L$_{\rm 60\mu m}$$<$12.06~L$_{\sun}$ (\cite{gmn88}), 
and Cygnus~A has log~L$_{\rm 60\mu m}$=11.72~L$_{\sun}$,
luminosities roughly an order of magnitude lower than those of the two RQQs.
The far-IR luminosity could still be produced by interaction-induced 
starbursts, but it could also be reprocessed AGN emission.  
Since it is impossible to disentangle the two, the far-IR luminosities for 
these two radio galaxies are inconclusive.
We note that CO(1-0) observations have been made of Cygnus~A 
(\cite{ba94,eva96}) and CO(3-2) observations of \9104\ (\cite{eva96}).
Neither object was detected, but it is unclear how to 
extrapolate these cold gas mass limits to total gas mass limits,
particularly if the gas is very near the central engine 
(or if it is immersed in a dense, hot cooling flow),
as Barvainis \& Antonucci point out.

\paragraph{Optical Properties:} \label{opt}
{\bf RQQs:} The host galaxy of \1821\ is a featureless red cD galaxy which is
slightly asymmetrical and offset from the nucleus by 1-2\arcsec\ (\cite{hn91b}).
Hutchings \& Neff (1991a) subjectively classify the galaxy as
undergoing a weak interaction of somewhat old age.
The host galaxy of \9104\ is a cD galaxy possibly in the midst of cannibalizing
several smaller galaxies (\cite{soi96}).
Hutchings \& Neff (1991a) subjectively classify it as
undergoing only a somewhat weak interaction of moderate age.
Thus the optical evidence for mergers or strong interactions in these two RQQ
host galaxies is suggestive but not strong.
{\bf RLQs:} A 600~sec archival WFPC2 image of 3C~206 shows strong evidence for
interaction of its elliptical host galaxy with several smaller galaxies.
The host galaxy's isophotes are slightly asymmetrically extended to the WNW, 
and two knots of emission, possibly galaxies being swallowed, appear within the
host galaxy 1.5\arcsec\ SE and 0.5\arcsec\ W of the quasar.
Projection effects cannot be ruled out,
but the chances of such close projections occurring are quite small.
A third galaxy, 4.25\arcsec\ SSW of the quasar, shows a faint nucleus inside a
distorted envelope of low surface brightness emission, consistent with being
tidally disrupted by the quasar host.
A 280~sec archival WFPC2 image of 3C~263 shows five faint knots of emission
within 5\arcsec\ of the quasar
and a very faint, possibly asymmetrical, underlying envelope of emission.
Better data are needed to determine the galaxy's morphology,
as the current data do not rule out e.g. a luminous spiral galaxy host
(which would however be unprecedented for a RLQ).
No information is available on the host galaxy of \2352.
{\bf PRGs:} An archival WFPC2 image of 3C~295 shows that its host galaxy is 
definitely disturbed, with distorted, asymmetrical isophotes and a partial
shell or plume of emission.
Optical evidence for interaction in Cygnus~A (\cite{sh89,srl94,j96})
is less obvious but still strong:
a secondary IR peak 1$\arcsec$\ north of the nucleus,
substantial dust in the inner regions of the galaxy, counter-rotating gas
structures and evidence for star formation in the nuclear region, 
and twisted isophotes (which might not indicate interaction, however;
cf. Smith and Heckman 1989).
Thus the 3C~295 host galaxy has almost certainly undergone a merger or strong
interaction which could have begun any time within the last $\sim$Gyr, and
Cygnus~A 
probably also has been disturbed (as suggested by Stockton, Ridgway \& Lilly
1994), but by a less disruptive or less recent event.

\bigskip

Thus while the 
evidence is not conclusive except in the case of 3C~295, it is on the whole 
supportive of a scenario where these AGN host galaxies have undergone strong 
interactions or mergers.  Specifically, {\it in no case where data is available
is no evidence for interaction seen.}  
This lends support to our suggestion that strong interactions may be the sole
mechanism needed to explain the presence and evolution of powerful AGN in clusters.

However, the observations do not rule out the validity of the cooling flow model
for at least some objects.  
While the cooling flow model need not be applicable to the host clusters of the
low-z AGN Cygnus~A, 3C~206, and \1821\ for it to explain the evolution of FR~II
AGN environments at z$\gtrsim$0.4, it should apply to the others.
In \9104\ and possibly 3C~295, the central ICM densities may reach the values
required by the cooling flow model without need for an additional mechanism.
But in any case, the z$\gtrsim$0.4 AGN host galaxies show evidence for having
undergone strong interaction(s) capable of boosting the central densities
sufficiently high for the Compton-cooling feedback mechanism to occur.
(Average densities within the central $\sim$100~pc of up to 2900~\cm3\ have been
inferred from CO observations of interacting or merging gas-rich galaxies (\cite{sco91}).)
Also, cooling flows may be preferentially located in more compact and
cooler clusters at z$\gtrsim$0.4, which would make their detection more 
difficult in our data.

In summary,
we suggest that strong interactions with gas-containing galaxies may be
the only mechanism needed to explain the presence and evolution of powerful
AGN in clusters.
This suggestion is consistent with the far-IR and optical properties of the host
galaxies of the AGN discussed in this paper, despite the rarity of such encounters in the high-\sv, high ICM density cluster environments of some of those AGN.
However, the cooling flow model cannot be ruled out for at least some objects.  
The data most needed to help determine the relative importance of 
each process 
are X-ray imaging, optical imaging, and \sv\ measurements of powerful AGN
host clusters.
The strong interaction scenario 
predicts that 
the host galaxies of all AGN in clusters should show signs of interaction,
and that the host clusters will rarely have high velocity dispersions, 
and rarely high X-ray luminosities and ICM densities as well.
The cooling flow model predicts that FR~II AGN host clusters at high z
should have cooling flows, but not necessarily at low z.

\section{Conclusions}	\label{finale}

We report a limit of 1.63~10$^{44}$~ergs~s$^{-1}$ on the rest-frame 0.1-2.4~keV
X-ray luminosity of the host cluster of the RLQ 3C~206
(assuming r$_{\rm core}$=125~kpc and kT=2.5~keV)
and a detection of \lx=3.74$\pm$0.57~10$^{44}$~ergs~s$^{-1}$ for the host
cluster of the RQQ \1821 (\ho=50, \qo=0.5 for both values).
CL~1821+643 is one of the most X-ray luminous clusters known, 
overluminous for its optical richness (also the case for \9104), and it has
a cooling flow of \mcool=1120$\pm$440~h$_{50}^{-2}$~M$_{\sun}$~yr$^{-1}$.
Its existence complicates interpretation of X-ray spectra of this
field (Appendix \ref{yaq}).  In particular, the observed
Fe~K$\alpha$ emission is probably solely due to the cluster,
and either the quasar is relatively X-ray quiet for its optical luminosity
or the cluster has a relatively low temperature for its luminosity.

We combine our data with the recent observation of X-ray emission from the 
host cluster of the buried RQQ \9104\ (\cite{fc95}), our previous upper limits
for two RLQs at z$\sim$0.7 (\cite{paper1}), and literature data on FR~II radio 
galaxies.  
We compare this dataset to the predictions of three simple models for the
presence and evolution of powerful AGN in clusters: 
the cooling flow model, the low-ICM-density model, and the low-\sv\ model.

In the cooling flow model (\S\ref{cf}),
FR~II AGN host clusters at z$\gtrsim$0.4 have dense cooling flows.  
Cooling flows have been detected in a few PRG and RQQ host
clusters (Cygnus~A, \1821, \9104, 3C~295).  
However, three RLQ host clusters (\2352, 3C~263, 3C~206)
have \mcool$\lesssim$200~\myr, unless cooling
flows are preferentially found in cooler, denser clusters at high z
or some mechanism besides the cooling flow has increased the central densities
in those clusters to create the high central pressures required by this model.
Strong interactions with gas-containing galaxies could be that mechanism.
Nevertheless, it is likely that the cooling flow model is not the {\it sole}
explanation for the presence and evolution of powerful AGN in clusters.

In the low-ICM-density model (\S\ref{icm}), FR~II AGN form in low-density ICM 
clusters and are destroyed as the ICM density increases.  
The three RLQs in our sample have host clusters consistent with this model, but
the two FR~II PRGs and the two RQQs have high-density host clusters overluminous
for their optical richnesses.
This is consistent with recent radio and X-ray studies of radio sources in 
different environments at z$\sim$0.5 which show no evidence for dense ICM in the
majority of powerful AGN host clusters at that redshift 
(\cite{rse95,hut96,wd96}), but not with radio work around z$\sim$1 PRGs which
infers gas densities and temperatures similar to nearby clusters (Wellman, Daly
\& Wan 1996ab, \cite{car97}), or X-ray work which detects extended emission with
\lx$\sim$10$^{44}$~erg~s$^{-1}$ around z$>$1 radio galaxies 
(\cite{sd95}, Crawford \& Fabian 1996a, 1996b).
These data show that the low-ICM-density model cannot be the only mechanism
behind the presence and evolution of powerful AGN in clusters.
Nonetheless, it is possible that most powerful AGN host clusters have low ICM
densities, and that the exceptions are either old AGN which originally formed
when the cluster ICM was less dense, or rare cases of strong interactions with
galaxies which retained some of their gas in high-ICM density environments.

In the low-\sv\ interaction/merger model (\S\ref{sigv}), FR~II AGN 
are preferentially found in clusters with low velocity dispersions, where the 
strong interactions which can create powerful AGN are more common.
Only a handful of \sv\ measurements for powerful AGN host clusters exist. 
The measurements of CL~3C~206 and the composite quasar host cluster of EYG91
support this model, those of CL~3C~295 and CL~1821+643 do not, and
CL~CygA lacks an accurate \sv\ determination.
More data are needed to be definitive.

We suggest that strong interactions with gas-containing galaxies may be
the only mechanism needed to explain the presence and evolution of powerful
AGN in clusters.
The far-IR and optical properties of the host galaxies of the AGN discussed in
this paper are consistent with this strong interaction scenario (\S\ref{farir}),
despite the rarity of such encounters in the high-\sv, high ICM density cluster
environments of some of those AGN.
However, the cooling flow model cannot be ruled out for at least some objects.
The relative importance of strong interactions and cooling flows 
can be determined by
testing the predictions of the models with future observations.
The cooling flow model predicts that FR~II AGN at z$\gtrsim$0.4 will be found
in dense cooling flow clusters and that if the cooling flows do not provide the 
necessary high central pressures for the Compton-cooling feedback mechanism to
work, there should be evidence for an additional process which has increased the
pressure, such as a strong interaction with a gas-containing galaxy.
The strong interaction scenario predicts that the host galaxies of all AGN in
clusters should show signs of interaction,
and that the host clusters will rarely have high velocity dispersions
or high X-ray luminosities and ICM densities.
Unlike the cooling flow model, the strong interaction scenario has the advantage
that it is applicable to FR~II AGN in all environments, not just clusters.

To definitively rule out some of the models we have considered and to advance
our understanding of the relationships between powerful AGN and their host 
clusters, the following data will be needed:  
1) more X-ray observations of FR~II AGN in clusters, especially those for which
extended emission-line regions have been studied by Bremer et~al. (1992) and 
others, to ascertain whether these AGN host clusters are more likely
to have cooling flows or low-density ICM ({\sl ROSAT} HRI data on 3C~215 and
3C~254 received after this paper was submitted do not show luminous cluster
X-ray emission);
2) accurate measurements of \sv\ (and \bgq\ where necessary) for FR~II AGN host
clusters, to test the low-\sv\ model;
3) more detailed studies and modelling of the host galaxy properties of 
AGN in clusters,
to rigorously test our strong interaction scenario for their origins;
4) optical and X-ray studies of the rare RQQs in rich clusters, to determine
the properties of their host clusters and why they are not radio-loud AGN;
5) searches for extended low-frequency emission from remnant radio lobes 
around \1821\ and \9104, to test the idea that these RQQs were once RLQs.
%

\acknowledgements

We thank Julio Navarro and the referee for their helpful comments
and the \ROSAT\ AO6 TAC for their support of this project and for pointing out
the existence of the archival \ein\ observation of 3C~206 to us.
This research has made use of 
data obtained through the High Energy Astrophysics Science Archive Research
Center Online Service, provided by the NASA-Goddard Space Flight Center;
data from the NRAO VLA Sky Survey obtained through the 
Astronomy Digital Image Library, a service of the
National Center for Supercomputing Applications;
data from operations made with the NASA/ESA Hubble Space Telescope, 
obtained from the data archive at the Space Telescope Science Institute, 
which is operated by the Association of Universities for Research in Astronomy,
Inc., under NASA contract NAS 5-26555;
and data from the IRAS archive at the Infrared Processing and Analysis Center 
and the NASA/IPAC Extragalactic Database (NED), which are operated by the Jet
Propulsion Laboratory, California Institute of Technology, under contract
to NASA.
PBH acknowledges support from an NSF Graduate Fellowship
and from NASA funding for analysis of {\sl ROSAT} observations.

\appendix
\section{X-ray Properties of Objects in the Field of \1821}	\label{yaq}

The field of \1821\ has been extensively monitored in the
optical (\cite{ul92,kol93}, hereafter K93), 
UV (\cite{kol91,ul92,lee93},K93),
EUV (\cite{fru95}),
and X-ray (\cite{pm84,wby89,kii91,wil92,ul92,yaq93}, K93, \cite{ys94,yam94}).
Confirmation of the existence of a luminous cooling flow cluster
surrounding the quasar complicates the interpretation of previous X-ray
observations of the field, since even the \rosat\ PSPC 
could not readily resolve the emission from CL~1821+643,
although it did resolve the white dwarf K1-16 from \1821.

To assist in future modelling, in Table 3
we present basic parameters for the different components as measured with
the \rosat\ HRI.
\begin{table}
\dummytable\label{tbl-3}
\end{table}
Column one gives the {\it total} HRI count rate for each component, calculated
from the fits made to the radial profiles extracted from the HRI image.
Converting from \ctsec\ to flux requires assuming an 
absorbing column density and spectrum for each component.
We assume a Galactic log~N$_{\rm H}$=20.58 (\cite{ls95}).
The \rosat\ HRI has extremely limited energy resolution, so the spectral
parameters of the different components cannot be computed from our data alone.
However, K93 have fit the combined spectra of the cluster and quasar using
\rosat\ PSPC data, whose energy response is more similar to the HRI than any
other instrument.
We use their spectral parameters for the quasar and white dwarf, 
but neglect the observed soft excess component of the quasar spectrum.
We obtain a total unabsorbed quasar plus cluster flux of 
3.97 10$^{-11}$ \cgsflux, compared to the 5.63 10$^{-11}$ \cgsflux\ found 
by K93 (calculated from \lx\ in their Table 1).
The difference can be entirely explained by the soft excess; 
attributing 10\% of our observed quasar flux to a blackbody soft excess 
(\cite{sax93}) with kT=0.04~keV would result in a flux equal to that of K93.

Although spectral fitting is really needed to do so,
we can use previous X-ray observations of this field 
in conjunction with our spatially resolved emission measurements
to obtain useful information about the cluster and quasar properties.
In the following discussion all luminosities are in the quasar {\it rest frame}
bandpasses and scale as h$_{50}^{-2}$.
All hard X-ray measurements since 1980 are consistent with 
\lx(2-10~keV) = 8.2~10$^{45}$ erg~s$^{-1}$ (\cite{yaq93}).  
With a cluster \lx(0.1-2.4~keV)=2.33$\pm$0.57~10$^{45}$ erg~s$^{-1}$
ignoring the cooling flow component,
we expect a cluster \lx(2-10~keV)=5.62 10$^{45}$~erg~s$^{-1}$
for a 10.5 keV thermal brehmsstrahlung (TB) spectrum.
If the quasar has a 2-10~keV to B-band flux ratio equal to the lowest observed
among the other eleven quasars in the sample of Williams et~al. (1992), 
then 34\% of the observed 2-10~keV emission would be from the quasar and 66\%
from the cluster, for a cluster \lx(2-10~keV)=5.41 10$^{45}$~erg~s$^{-1}$,
consistent with expectations for a 10.5 keV TB spectrum.
Using a Raymond-Smith spectrum to calculate the bandpass correction
should not change these results drastically.
A possible problem with a single-temperature TB spectrum is that the 2-10~keV
(and indeed the 2-18~keV) spectrum is consistent with a smooth power law,
the only significant residual feature being a probable 6.6~keV Fe K$\alpha$
line redshifted to 5.1~keV (\cite{kii91,yaq93,ys94,yam94}).
If 66\% of the 2-10~keV flux was from a 10.5~keV TB spectrum and 34\% from a
power law, we might expect a spectral break or residual feature near 8.1~keV
(observed).
A detailed fit is needed to determine how strong such a residual would be;
its apparent absence might simply be due to insufficient S/N in existing 
spectra (Yaqoob 1996, personal communication),
or might indicate a multiphase (non-isothermal) cluster whose integrated
spectrum better resembles a power law.
A range of temperatures is in fact expected for the cooling flow emission
component (i.e. the emission from gas cooling from the dominant cluster
temperature kT), which we have neglected so far.
Roughly speaking, we expect k\={T$_{\rm CF}$}=0.5kT (\cite{joh92}),
using a cooling function $\Lambda$(T)$\propto$T$^{1/2}$ (\cite{sar88}).
Energetically, for kT$\sim$7~keV and k\={T$_{\rm CF}$}$\sim$3.5~keV,
we can self-consistently ascribe 66\% of the 2-10~keV emission to the cluster
and cooling flow component.
This kT is slightly low but is still within the range expected for the 
cluster's \lx\ and $\sigma_v$.
In either case, the Fe K$\alpha$ line of EW$\sim$140~eV observed by ASCA
(\cite{yam94}) can probably be explained by cluster emission alone.

We qualitatively conclude that the 2-10~keV spectrum of this field can be 
consistent with kT=10.5~keV cluster emission, 
provided that the quasar is somewhat X-ray quiet.
More likely, there is a range of temperatures in the cluster and 
the average temperature is $<$10.5~keV.
This is particularly true if the quasar is more normal in its X-ray/optical
properties, since attributing more 2-10~keV flux to the quasar translates
directly into a lower average temperature for the cluster.  To make these
constraints truly quantitative requires refitting existing X-ray spectra while
incorporating our spatially resolved emission constraints in the HRI band.



\begin{deluxetable}{llllll}
\tablenum{1}
\scriptsize
\tablewidth{0pc}
\tablecaption{Summary of Target Observations and Properties}
\tablehead{
\colhead{} & \colhead{3C~206} & \colhead{3C~263} & \colhead{PKS~2352-342} & \colhead{H~1821+643} & \colhead{IRAS~09104+4109}} 
\startdata
RA~(1950.) &	08:37:27.95   	& 11:37:09.34 	& 23:52:50.62 	& 18:21:41.89 	& 09:10:32.84 \nl 
Dec~(1950.) &	--12:03:54.2 	& +66:04:26.9 	& --34:14:39.5 	& +64:19:01.05 	& +41:08:53.61 \nl 
N$_{\rm H}$,~10$^{20}$~cm$^{-2}$ & 5.62	& 0.91	& 1.07 		& 3.80 		& 1.60 \nl 	
Observation & 	11/16-19/1979 & 11/4-7/1991 & 5/18-6/13/1993 & 3/22/1994 & 11/8/94 \nl 
~~~Dates & 	\nodata 	& 4/18-21/1993 & \nodata	& 10/20-24/1995 & \nodata \nl 	
Livetime,~sec &	
62732.75 & 26036.41 & 40173.94 & 30591.15 & 7937 \nl 
Redshift &
0.1976 	 & 0.646    & 0.706    & 0.297 	& 0.442 \nl 
D$_{\rm Luminosity}$, 10$^{27}$ cm &
3.82 	 & 13.4      & 14.8    	& 5.81    & 8.93 \nl 	
\cutinhead{Quasar Properties}
M$_{\rm B}$ 	&
-25.0	& -26.6	& -26.7	& -27.2	& -26.3$\pm$0.9\tablenotemark{a} \nl 
log~(L$_{\rm 60\mu m}$/L$_{\sun}$) & 
11.18	& $<$12.28 & (12.4)\tablenotemark{b} & 13.0 & 13.2 \nl 
log~P$_{\rm 20cm}$,~W~Hz$^{-1}$ &
26.45 	& 27.30  & 27.21  & 25.13   & 25.16 \nl 	
Radio~Morphology &
FR~II 	& FR~II & \nodata & core+lobe? & FR~I/II \nl 	
LLS\tablenotemark{c},~kpc 	&
799 	& 342   & (200)\tablenotemark{d} & 20  & 175 \nl 	
M$_{\rm host~galaxy}$ &
-22.6 (r) & $>$-23.2 (B) & \nodata & -24.1 (R) & -24.45\tablenotemark{e} (R) \nl 
\cutinhead{Cluster Properties}
B$_{\rm gq}$, Mpc$^{-1.77}$ &	683$\pm$197 	& 993$\pm$550 	& 681$\pm$280 	& 1200$\pm$200 	& 1210$\pm$293 \nl 
$\sigma_v$, km~s$^{-1}$ &
500$\pm$110 	& \nodata & \nodata & 1046$\pm$108  & \nodata \nl 	
r$_{\rm core}$, kpc &
(125)	& (125) & (125) & 176.8$\pm$9.9	& 200 \nl 	
L$_{\rm X,44}$\tablenotemark{f} &
$<$1.63	& $<$3.48 & $<$3.09 & 37.4$\pm$5.7 & 30.3$\pm$2.6 \nl 	
n$_{\rm e,0}$\tablenotemark{g}, 10$^{-3}$ cm$^{-3}$ &
$<$6.76 (7.01) & $<$7.34 (10.0) & $<$6.79 (9.56) & $>$81$\pm$22 & $>$(27$-$97) \nl 
T$_{\rm cool}$\tablenotemark{g}, Gyr &
$>$6.8 (6.5)    & $>$8.8 (6.5)  & $>$9.5 (6.8)  & $<$6.37$\pm$1.24    & $<$(3.65$-$1.01) \nl 
L$_{\rm X,44;~cooling~flow}$\tablenotemark{g} &
$<$0.82 (0.90) & 0 ($<$2.42) & 0 ($<$2.21) & 14.1$\pm$4.7 & 9.09$\pm$0.78 \nl 
\.{M}$_{\rm cool}$\tablenotemark{g}, M$_{\sun}$~yr$^{-1}$ &
$<$137 (150) & 0 ($<$202) & 0 ($<$184) & 1120$\pm$440 & 1003$^{+202}_{-272}$ \nl 
P$_{\rm central}$\tablenotemark{g}, 10$^6$ cm$^{-3}$K &
$<$0.20 (0.20) & $<$0.43 (0.58) & $<$0.40 (0.55) & $>$9.9$\pm$3.4 & $>$(3.6$-$12.8) \nl 
\tablebreak
\phm{D$_{\rm Luminosity}$, 10$^{27}$ cm} & \phm{11/16-19/1979} & \phm{4/18-21/1993} & \phm{5/18-6/13/1993} & \phm{10/20-24/1995} & \phm{11/8/94} \nl 
\enddata
\tablecomments{H$_{\rm o}$=50 and q$_{\rm o}$=\case{1}{2} assumed, although
values of n$_{\rm e}$, T$_{\rm cool}$, L$_{\rm X,44;~cooling~flow}$, and
\.{M}$_{\rm cool}$ are also given in parentheses for H$_{\rm o}$=50 and
q$_{\rm o}$=0.  Luminosities scale as h$_{50}^{-2}$, sizes as h$_{50}^{-1}$,
n$_{\rm e}$ as h$_{50}^{1/2}$, and T$_{\rm cool}$ as h$_{50}^{1/2}$.}
\tablenotetext{a}{Unobscured estimate from Hines \& Wills (1993), converted to
H$_{\rm o}$=50, q$_{\rm o}$=0.5.}
\tablenotetext{b}{2.4-2.9$\sigma$ IRAS SCANPI detection. 
Can be considered a 3$\sigma$ upper limit.}
\tablenotetext{c}{Largest linear size of the radio source, measured at 20 cm
except for 3C~206 (6 cm).}
\tablenotetext{d}{Very uncertain estimate.  The NVSS (cf. Condon et~al. 1994)
shows PKS~2352-342 to be slightly resolved (52\farcs6 vs. 46\arcsec\ for a source
6\arcmin\ away).  Modeling it as a gaussian yields the FWHM=200 kpc given here.}
\tablenotetext{e}{Contaminated by strong narrow line emission.}
\tablenotetext{f}{Rest-frame 0.1-2.4~keV luminosity in units of 10$^{44}$
erg~s$^{-1}$.}
\tablenotetext{g}{Values for q$_{\rm o}$=0 are in parentheses after values for
q$_{\rm o}$=0.5, both assuming H$_{\rm o}$=50.
Total estimated central density includes cooling flow component;
central densities for the King component only are 14.7$\pm$2.4 and 6.9$\pm$1.4
10$^{-3}$ cm$^{-3}$ for H~1821+643 and IRAS~09104+4109 respectively.
Central pressures calculated assuming kT=2.5~keV for 3C~206, kT=5~keV for 3C~263
and PKS~2352-342, kT=10.5$\pm$2.2 for H~1821+643, and kT=11.4 for IRAS~09104+4109.}
\tablerefs{
N$_{\rm H}$: Lockman \& Savage (1995) for all except 3C~206: Elvis, Lockman \&
Wilkes (1989) and IRAS~09104+4109: Fabian et~al. (1994).
L$_{\rm 60\mu m}$: 3C~206 and 3C~263: Neugebauer et~al. (1986); 
PKS~2352-342: this work; RQQs: Hutchings \& Neff (1991a).
Radio: 3C~206: Miley \& Hartsuijker (1978); 3C~263: Hutchings et~al. (1996);
PKS~2352-342: Quinento \& Cersosimo (1993) and Condon et~al. (1994);
H~1821+643: Kolman et~al. (1993) and Blundell \& Lacy (1995);
IRAS~09104+4109: Hines \& Wills (1993).
M$_{\rm host~galaxy}$: 
3C~206: Ellingson et~al. (1989); 3C~263: Crawford et~al. (1991);
H~1821+643: Hutchings \& Neff (1991b); IRAS~09104+4109: Kleinmann et~al. (1988).
B$_{\rm gq}$: 3C~206: Ellingson et~al. (1989); 3C~263, PKS~2352-342: Yee \& 
Ellingson (1993); H~1821+643: Lacy, Rawlings \& Hill (1992); IRAS~09104+4109: 
this work.
$\sigma_v$: 3C~206: Ellingson et~al. (1989); H~1821+643: this work.
Cluster X-ray properties: this work or Paper~I, except r$_{\rm core}$ and 
L$_{\rm X,44}$ (0.1-2.4~keV, observed) for IRAS~09104+4109, from
Fabian \& Crawford (1995).
}
\end{deluxetable}

\clearpage

\begin{deluxetable}{lcccccccccc}
\scriptsize
\tablenum{2}
\tablewidth{0pc}
\tablecaption{Quasar Host Cluster X-Ray Luminosities --- Upper Limits and Detections}
\tablehead{
\colhead{Core Radius} &
\colhead{125 kpc} &
\colhead{125 kpc} &
\colhead{125 kpc} &
\colhead{125 kpc} &
\colhead{250 kpc} &
\colhead{250 kpc} &
\colhead{250 kpc} &
\colhead{250 kpc} \\[.2ex]
\colhead{H$_{\rm o}$, q$_{\rm o}$} &
\colhead{50, 0.5} &
\colhead{75, 0.5} &
\colhead{50, 0.0} &
\colhead{75, 0.0} &
\colhead{50, 0.5} &
\colhead{75, 0.5} &
\colhead{50, 0.0} &
\colhead{75, 0.0}}
\startdata
3C~206 & 1.63 & 0.92 & 1.80 & 1.02 & 2.60 & 1.57 & 2.88 & 1.74 \\
3C~263\tablenotemark{a} & 3.48 & 1.86 & 4.83 & 2.58 & 5.10 & 3.26 & 7.08 & 4.52 \\
PKS~2352-342\tablenotemark{a} & 3.09 & 1.66 & 4.41 & 2.37 & 4.44 & 2.75 & 6.32 & 3.92 \\
H~1821+643\tablenotemark{b} & 51.5$\pm$10. & 22.9$\pm$4.4 & 44.8$\pm$8.8 & 19.9$\pm$3.9 & \nodata & \nodata & \nodata & \nodata \\
IRAS~09104+4109\tablenotemark{c} & \nodata & \nodata & \nodata & \nodata & 30.3$\pm$2.6 & 13.5$\pm$1.2 & 38.0$\pm$3.2 & 16.9$\pm$1.4 \\
\enddata
\tablecomments{Luminosities are in units of 10$^{44}$ ergs s$^{-1}$ in the
{\it rest-frame} 0.1--2.4~keV passband.  Values with uncertainties are 
detections; all other values are 3$\sigma$ upper limits.}
\tablenotetext{a}{These values are slightly larger than those of Table 2 
in Paper~I because they have been corrected to the rest-frame 0.1--2.4~keV 
passband.}
\tablenotetext{b}{We find a best-fit 
r$_{\rm core}$=176.8$\pm$9.9~h$_{50}^{-1}$~kpc 
(190.5$\pm$9.6~h$_{50}^{-1}$~kpc) for q$_{\rm o}$=0.5 (0).}
\tablenotetext{c}{Fabian \& Crawford (1995) find r$_{\rm core}$=30\arcsec,
or 200 (225) kpc for q$_{\rm o}$=0.5 (0), and a large cooling flow excess.}
\end{deluxetable}


\begin{deluxetable}{lcccccccc}
\scriptsize
\tablenum{3}
\tablewidth{0pc}
\tablecaption{Properties of Objects in the Field of H~1821+643}
\tablehead{
\colhead{Component} &
\colhead{HRI counts/s} &
\colhead{Central S.B.\tablenotemark{a}} &
\colhead{r$_{\rm core}$/$\sigma_{\rm gauss}$} &
\colhead{Spectrum\tablenotemark{b}} &
\colhead{ECF\tablenotemark{c}} &
\colhead{Unabsorbed Flux\tablenotemark{d}}}
\startdata
Quasar	     & 0.3136$\pm$0.0032		& \nodata 		    & \nodata & PL, $\alpha$=1.81 & 	      0.100 & 3.136$\pm$0.032 \nl
Cluster\tablenotemark{e} & 0.1152$\pm$0.0163	& 1.77$\pm$0.15 & 32\farcs2$\pm$1\farcs8 & RS, kT=5 keV &   0.223 & 0.516$\pm$0.073 \nl
Cooling Flow\tablenotemark{f} & 0.0699$\pm$0.0233 & 36.03$\pm$11.63 & 5\farcs56$\pm$0\farcs46 & RS, kT=5 keV & 0.223 & 0.314$\pm$0.105 \nl 
\nodata      & \nodata				& 43.85$\pm$18.21 & 5\farcs04$\pm$0\farcs51 & \nodata &	    \nodata & \nodata \nl
White Dwarf  & 0.0184$\pm$0.0011		& \nodata 		    & \nodata &	BB, kT=0.01	&	    0.00195 & 9.44$\pm$0.57 \nl
\enddata
\tablenotetext{a}{Central surface brightness in units of 10$^{-5}$ {\sl ROSAT}
HRI counts s$^{-1}$ arcsec$^{-2}$.}
\tablenotetext{b}{Spectrum assumed to convert counts to flux: PL=Power Law,
RS=Raymond-Smith plasma, BB=blackbody.}
\tablenotetext{c}{Energy to Counts conversion Factor, estimated from David et~al. (1995).}
\tablenotetext{d}{Observed 0.1-2.4 keV (rest-frame 0.13-3.11 keV) flux in units
of 10$^{-11}$ ergs s$^{-1}$ cm$^{-2}$,
corrected for Galactic absorption of log N$_{\rm H}$=20.58 only.}
\tablenotetext{e}{King model component only.  The central surface brightness
has been corrected by +1.9\% and r$_{\rm core}$ by +3.6\% to account for
systematic errors in the fitting as measured from simulated data.}
\tablenotetext{f}{Upper row values are not corrected for convolution with
{\sl ROSAT} HRI PRF; lower row values are corrected.}
\end{deluxetable}





\begin{figure}
\plotone{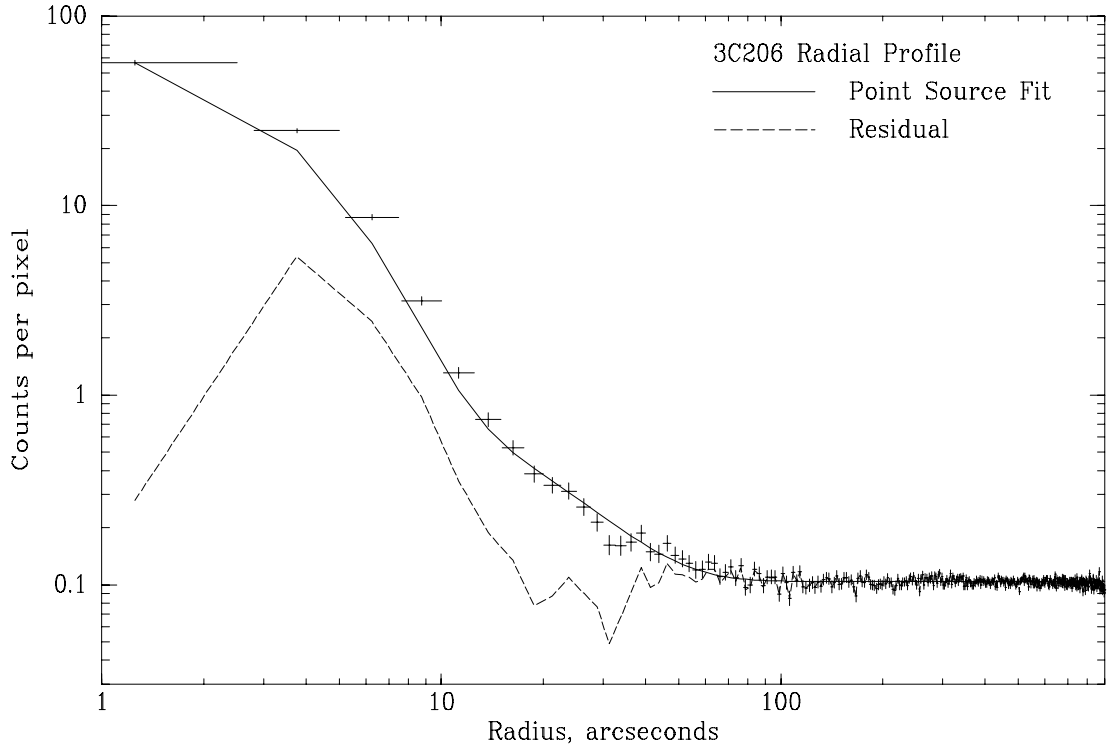}
\caption{
Radial profiles of the {\sl EINSTEIN} HRI image of the quasar 3C~206, 
the fitted PRF and background, and the residual  
after subtraction of an \ein\ HRI point response function (PRF) 
normalized such that the residual is zero in the innermost bin.
The residual is exaggerated in this log-log plot; note that the apparent
excess emission is of the same scale as the PRF and that the
background-subtracted residual is actually negative between 15-40\arcsec,
exactly the region in which cluster emission should be most prominent.
}
\label{206fit}
\end{figure}

\begin{figure}
\plotone{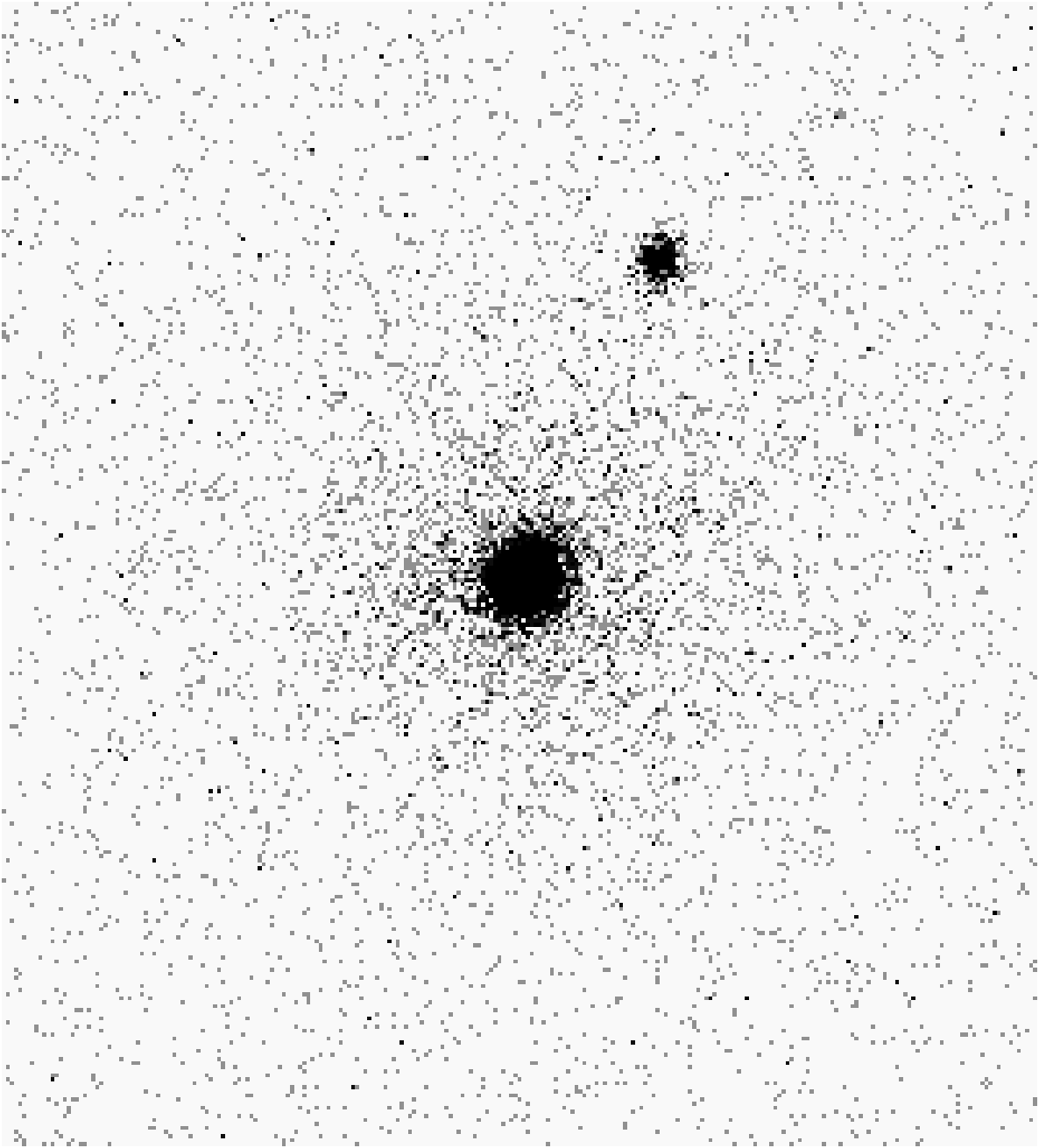}
\caption{
The central portion of the \rosat\ HRI image of the \1821\ field.
North is up and east is to the left.
The image has been binned into 1\arcsec\ pixels but has not been smoothed.
The quasar is at the center of the image and the white dwarf is 88\arcsec\ to
the NW.
The quasar is clearly surrounded by excess emission, above that expected from
a point source, from its host cluster.
}
\label{1821img}
\end{figure}

\begin{figure}
\plotone{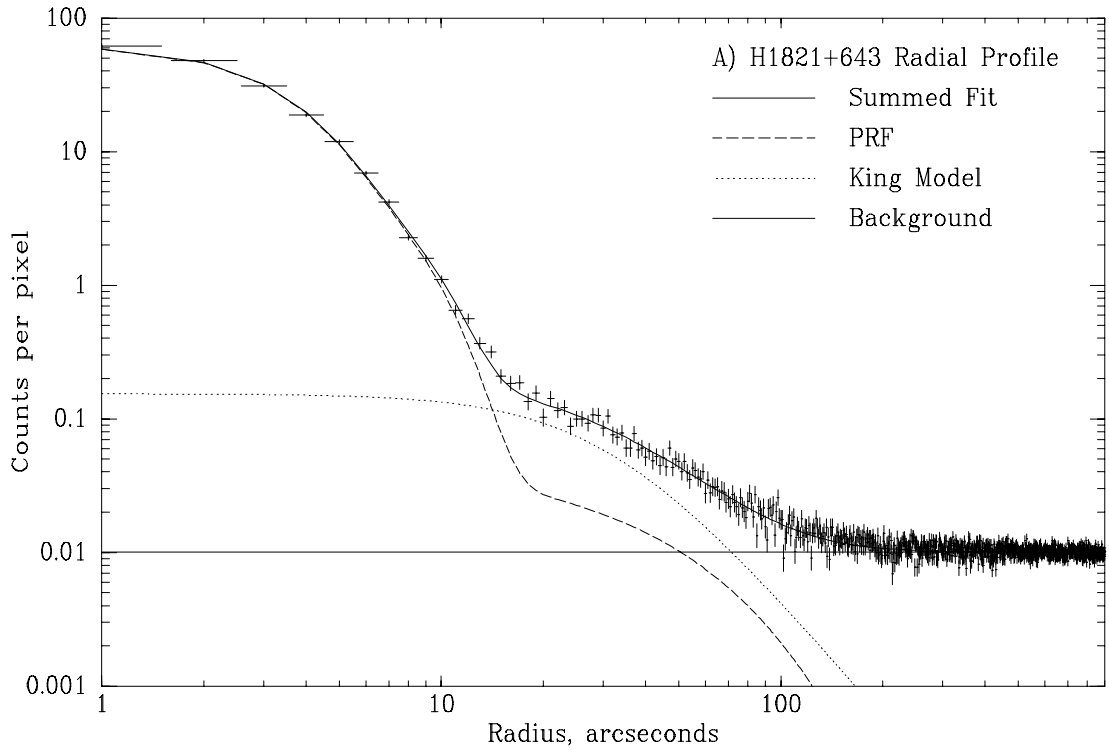} \\
\plotone{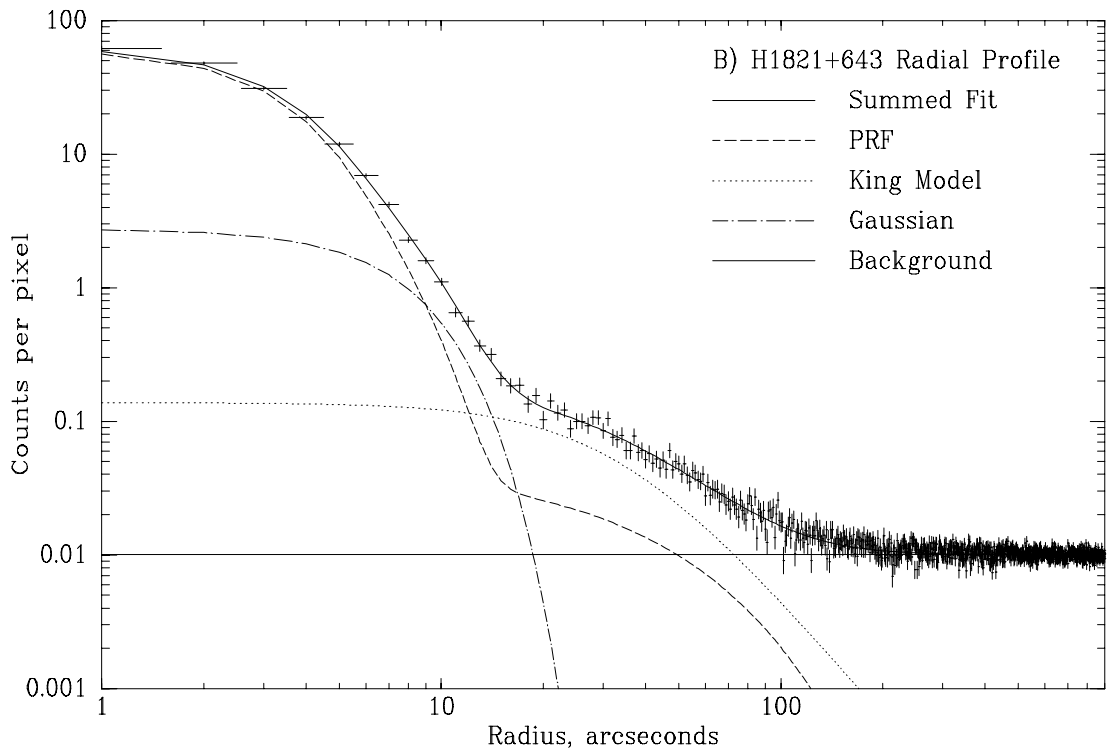}
\caption{
Observed radial profile of the quasar \1821\ and two fits to it.
a. Fit (solid line) includes background (straight solid line), PRF (dashed),
and King model (dotted).
b. Fit also includes gaussian component (dot-dash).
}
\label{1821fit}
\end{figure}

\begin{figure}
\plotone{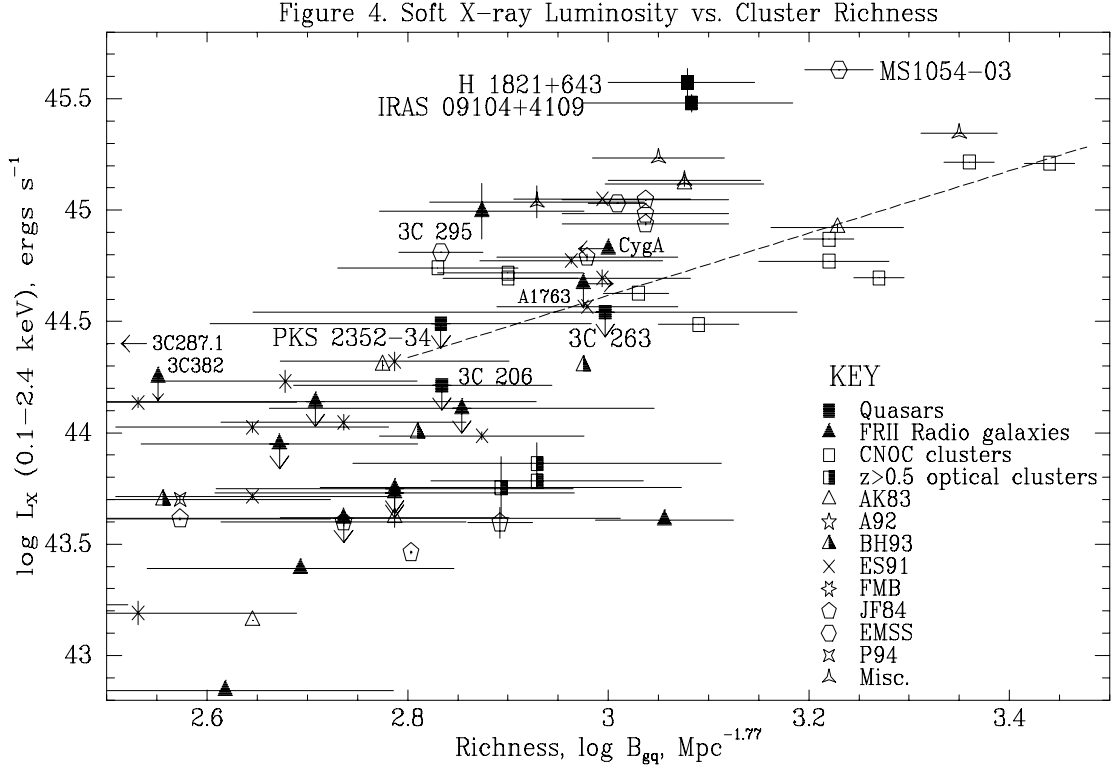}
\caption{
The amplitude of the galaxy-cluster center spatial correlation function
B$_{\rm gc}$ vs. rest-frame 0.1-2.4~keV X-ray luminosity L$_{\rm X}$.
Quasar host clusters (this work, Paper~I, and Fabian \& Crawford (1995))
are plotted as filled squares (upper limits assume r$_{\rm core}$=125~kpc).
Open squares are a \={z}$\sim$0.3 subsample of X-ray
selected EMSS clusters being studied by the CNOC group (Carlberg et~al. 1996),
filled triangles are FR~II PRG host clusters,
and other symbols are objects from the literature, as detailed in the key.
The dotted line is the best fit to the CNOC/EMSS data.
Half-filled triangles are typical Abell richness 0, 1, and 2 clusters with
\bgq\ values from EYG91 and \lx\ values from Briel \& Henry (1993).  These 
\lx\ values agree with those from Burg et~al. (1994) and Wan \& Daly (1996).
The typical luminosity for Abell richness 2 clusters falls below the average
of objects in our literature dataset, illustrating that said data are 
extremely inhomogeneous and should be viewed merely as illustrating the 
range of values observed.  
L$_{\rm X}$ references:
Quasars: this work, Paper~I, and Fabian \& Crawford (1995).
PRGs: Henry \& Henriksen (1986), Worrall et~al. (1994),
Crawford \& Fabian (1993,1995), O'Dea et~al. (1996), and Wan \& Daly (1996,
and references therein).
CNOC/EMSS clusters: Carlberg et~al. (1996).
z$>$0.5 optically selected clusters: Castander et~al. (1994),
Nichol et~al. (1994), and Roche et~al. (1995).
Literature reference keys:
AK83: Abramopoulos \& Ku (1983);
A92: Allen et~al. (1992);
BH93: Briel \& Henry (1993);
ES91: Edge \& Stewart (1991a,b);
FMB: Fabricant, McClintock \& Bautz (1991) and Fabricant, Bautz \& McClintock (1994);
JF84: Jones \& Forman (1984);
EMSS: Gioia \& Luppino (1994), Luppino \& Gioia (1995), and Nesci, Perola \& Wolter (1994);
P94: Pierre et~al. (1994a,b);
Misc.: Donahue \& Stocke (1995); Elbaz, Arnaud \& Boehringer (1995); 
Edge et~al. (1994ab); Hughes, Birkinshaw \& Huchra (1995); Schwartz et~al.
(1991); Smail et~al. (1995); Schindler et~al. (1996); and White et~al. (1994).
\bgq\ values taken from (or calculated from N$_{\rm 0.5}$ values given in)
Longair \& Seldner (1979), Bahcall (1981), Mathieu \& Spinrad (1981),
Prestage \& Peacock (1988), Yates, Miller \& Peacock (1989), Hill \& Lilly
(1991), Allington-Smith et~al. (1993), and Yee \& Ellingson (1993).
}
\label{LvB}
\end{figure}

\begin{figure}
\plotone{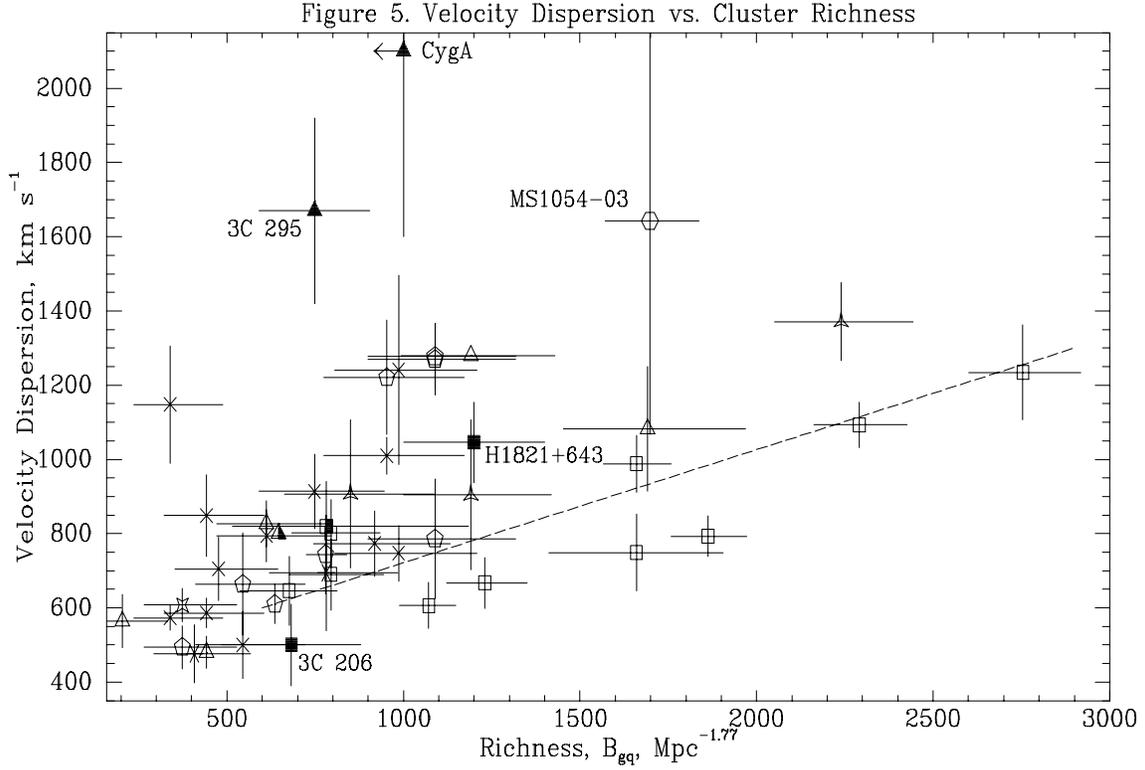}
\caption{
Cluster velocity dispersion versus richness.  Symbols as in Figure \ref{LvB}.
The half-filled triangle is a `typical' Abell richness 1 cluster using 
\bgq\ from EYG91 and $\sigma_v$ from Ellingson, Green \& Yee (1991).
The half-filled square is CL1322+3027 at z=0.757 (Castander et~al. 1994).
Velocity dispersions for other literature objects from
Struble \& Rood (1987),
Ellingson et~al. (1989),
Zabludoff, Huchra \& Geller (1990),
Teague, Carter \& Gray (1990),
and Edge \& Stewart (1991b).
\bgq\ references are listed in the legend to Figure \ref{LvB}.
}
\label{DvB}
\end{figure}


\scriptsize

\begin{thebibliography}{}

\bibitem[Aarseth \& Fall 1980]{af80} \reference{} Aarseth, S.J., and Fall, 
	S.M.  1980, \apj\ 236, 43


\bibitem[Abramopoulos \& Ku 1983]{ab83} \reference{} Abramopoulos, F., and Ku,
	W. H.-M.  1983, \apj\ 271, 446

\bibitem[Allen 1995]{all95} \reference{} Allen, S.W.  1995, \mnras\ 276, 947

\bibitem[Allen et~al. 1992]{all92} \reference{} Allen, S.W., Edge, A.C., 
	Fabian, A.C., Boehringer, H., Ebeling, H., Johnstone, R.M., Naylor, T.,
	and Schwarz, R.A.  1992, \mnras\ 259, 67

\bibitem[Allen et~al. 1995]{aea95} \reference{} Allen, S.W., Fabian, A.C., 
	Edge, A.C., Boehringer, H., and White, D.A.  1995, \mnras\ 275, 741


\bibitem[Allington-Smith et~al. 1993]{as93} \reference{} Allington-Smith, J.R.,
	Ellis, R.S., Zirbel, E.L., and Oemler, Jr., A.O.  1993, \apj\ 404, 521

\bibitem[Antonucci 1993]{ant93} \reference{} Antonucci, R.  1993, \araa\ 31, 473

\bibitem[Arnaud 1984]{arn84} \reference{} Arnaud, K.~A., Fabian, A.C., Eales,
	S.A., Jones, C., and Forman, W. 1984, \mnras\ 211, 981

\bibitem[Arnaud 1988]{arn88} \reference{} Arnaud, K.~A. 1988, in {\it Cooling 
     Flows in Clusters and Galaxies}, ed. A.C. Fabian (Dordrecht: Kluwer)

\bibitem[Bahcall 1981]{bah81} \reference{} Bahcall, N.~A. 1981, \apj\ 247, 787


\bibitem[Barthel \& Arnaud 1996]{ba96}\reference{} Barthel, P. D., and Arnaud,
     K.  1996, \mnras\ 283, L45

\bibitem[Barthel \& Miley 1988]{bm88}\reference{} Barthel, P. D., and Miley, 
     G. K. 1988, Nature 333, 319

\bibitem[Barvainis \& Antonucci 1994]{ba94} \reference{} Barvainis, R., and
	Antonucci, R.  1994, \aj\  107, 1291


\bibitem[Bechtold et~al. 1994]{bec94} \reference{} Bechtold, J., Elvis, M., 
	Fiore, F., Kuhn, O., Cutri, R.M, McDowell, J.C., Rieke, M., 
	Siemiginowska, A., and Wilkes, B.J.  1994, \aj\ 108, 759


\bibitem[Birkinshaw \& Worrall 1993]{bw93} \reference{} Birkinshaw, M., and
	Worrall, D.M.  1993, \apj\ 412, 568

%

\bibitem[Blandford 1996]{bla96} \reference{} Blandford, R.  1996, in {\it
	Cygnus~A -- Study of a Radio Galaxy}, eds. C.L. Carilli and D.E. Harris
	(Cambridge: Cambridge)

\bibitem[Blundell \& Lacy 1995]{bl95} \reference{} Blundell, K.M., and Lacy, M.
      1995, \mnras\ 274, L9

\bibitem[Blundell et~al. 1996]{blu96} \reference{} Blundell, K.M., Beasley,
	A. J., Lacy, M., and Garrington, S.T.  1996, \apjl, in press
	(astro-ph/9606102)

\bibitem[Boehringer et~al. 1993]{boe93} \reference{} Boehringer, H., Voges, W.,
	Fabian, A.C., Edge, A.C., and Newmann, D.M.  1993, \mnras\ 264, L25

\bibitem[Bower et~al. 1994]{bow94} \reference{} Bower, R.G., Boehringer, H.,
     Briel, U.G., Ellis, R.S., Castander, F.J., and Couch, W.J. 1994, 
     \mnras\ 268, 345

\bibitem[Bregman, McNamara \& O'Connell 1990]{bmo90} \reference{} Bregman, 
	J.N., McNamara, B.R., and O'Connell, R.W.  1990, \apj\ 351, 406


\bibitem[Bremer et~al. 1992]{brem92} \reference{} Bremer, M.N., Fabian, A.C.,
	Crawford, C.S., and Johnstone, R.M.  1992, \mnras\ 254, 614

\bibitem[Bremer, Fabian \& Crawford 1996]{bfc96} \reference{} Bremer, M.N.,
	Fabian, A.C., and Crawford, C.S.  1996, in {\it Cold Gas at High 
	Redshift}, eds. Bremer et~al. (Dordrecht: Kluwer)

\bibitem[Briel \& Henry 1993]{bh93} \reference{} Briel, U.G., and Henry, J.P.
	1993, \aap\ 278, 379

\bibitem[Burg et~al. 1994]{bur94} \reference{} Burg, R., Giacconi, R., Forman,
	W., and Jones, C.  1994, \apj\ 422, 37

\bibitem[Burns 1990]{b90} \reference{} Burns, J.O.  1990, \aj\ 99, 14



\bibitem[Canizares, Fabbiano \& Trinchieri 1987]{cft87} \reference{}
	Canizares, C.R., Fabbiano, G., and Trinchieri, G.  1987, \apj\ 312, 503

\bibitem[Carilli et~al. 1991]{car91} \reference{} Carilli, C.L., Perley, R.A.,
	Dreher, J.W., and Leahy, J.P.  1991, \apj\ 383, 554

\bibitem[Carilli, Perley \& Harris 1994]{cph94} \reference{} Carilli, C.L., 
	Perley, R.A., and Harris, D.E. 1994, \mnras\ 270, 173


\bibitem[Carilli et~al. 1997]{car97} \reference{} Carilli, C.L., R\"ottgering,
	H.J.A., van Ojik, R., Miley, G.K., and van Breugel, W.J.M.  1997,
	\apjs\ in press

\bibitem[Carlberg et~al. 1996]{car96} \reference{} Carlberg, R.G., Yee, H.K.C.,
	Ellingson, E., Abraham, R., Gravel, P., Morris, S., and Pritchet, 
	C.  1996, \apj\ 462, 32

\bibitem[Castander et~al. 1994]{cas94} \reference{} Castander, F.J., Ellis, 
     R.S., Frenk, C.S., Dressler, A., and Gunn, J.E. 1994, \apjl\ 424, L79

\bibitem[Cavaliere \& Padovani 1988]{cp88} \reference{} Cavaliere, A., and
	Padovani, P.  1988, \apj\ 315, 411


\bibitem[Cen \& Ostriker 1994]{co94} \reference{} Cen, R., and Ostriker, J.P.
	1994, \apj\ 429, 4

%
%

\bibitem[Condon et~al. 1994]{nvss} \reference{} Condon, J.J., Cotton, W.D.,
	Greisen, E.W., Yin, Q.F., Perley, R.A., and Broderick, J.J., 1994,
	preprint (http://www.cv.nrao.edu/$\sim$jcondon/nvss.html)

%

\bibitem[Crawford et~al. 1991]{cra91} \reference{} Crawford, C.S., Fabian,
	A.C., George, I.M., and Naylor, T.  1991, \mnras\ 248, 139

\bibitem[Crawford \& Fabian 1989]{cf89} \reference{} Crawford, C.S., and
	Fabian, A.C. 1993, \mnras\ 239, 219

\bibitem[Crawford \& Fabian 1993]{cf93} \reference{} Crawford, C.S., and
	Fabian, A.C. 1993, \mnras\ 260, L15

\bibitem[Crawford \& Fabian 1995]{cf95} \reference{} Crawford, C.S., and
	Fabian, A.C. 1995, \mnras\ 273, 827

\bibitem[Crawford \& Fabian 1996a]{cf96a} \reference{} Crawford, C.S., and
	Fabian, A.C. 1996a, \mnras\ 281, L5	

\bibitem[Crawford \& Fabian 1996b]{cf96b} \reference{} Crawford, C.S., and
	Fabian, A.C. 1996b, \mnras\ 282, 1483

\bibitem[Crawford \& Vanderriest 1996]{cv96} \reference{} Crawford, C.S., and
	Vanderreist, C. 1996, \mnras\ 281, in press


\bibitem[David et~al. 1995]{hri95} \reference{} David, L.P., Harnden, F.R., Jr.,
     Kearns, K.E., and Zombeck, M.V. 1995, The \ROSAT\ High Resolution Imager
     (HRI) (US {\sl ROSAT} Science Data Center/SAO) (D95)


\bibitem[Donahue \& Stocke 1995]{ds95} \reference{} Donahue, M., and Stocke,
	J. T.  1995, \apj\ 449, 554


\bibitem[DeRobertis \& Yee 1990]{dry90} \reference{} DeRobertis, M., and Yee, 
     H.~K.~C.  1990, \aj\ 100, 84

\bibitem[Dressler \& Gunn 1992]{dg92} \reference{} Dressler, A., and Gunn, J.E.
	1994, \apjs\ 78, 1

\bibitem[Duc \& Mirabel 1994]{dm94} \reference{} Duc, P.-A., and Mirabel, I.F.
	1994, \aap\ 289, 83


\bibitem[Ebeling et~al. 1996]{ebe96} \reference{} Ebeling, H., Voges, W.,
	Boehringer, H., Edge, A.C., Huchra, J.P., and Briel, U.G.  1996,
	\mnras, in press (astro-ph/9602080)

\bibitem[Edge \& Stewart 1991a]{es91a} \reference{} Edge, A.C., and Stewart,
	G.C.  1991a, \mnras\ 252, 414

\bibitem[Edge \& Stewart 1991b]{es91b} \reference{} Edge, A.C., and Stewart,
	G.C.  1991b, \mnras\ 252, 428


\bibitem[Edge et~al. 1994a]{zw3146} \reference{} Edge, A.C., Fabian, A.C.,
	Allen, S.W., Crawford, C.S., White, D.A., Boehringer, H., and Voges,
	W.  1994a, \mnras\ 270, L1

\bibitem[Edge et~al. 1994b]{s295} \reference{} Edge, A.C., Boehringer, H.,
	Guzzo, L., Collins, C.A., Neumann, D., Chincarini, G., De Grandi, S.,
	Duemmler, R., Ebeling, H., Schindler, S., Weitter, W., Vettolani, P.,
	Briel, U., Cruddace, R., Gruber, R., Gursky, H., Hartner, G.,
	MacGillivray, H.T., Schuecker, P., Shaver, P., Voges, W., Wallin, J.,
	Wolter, A., and Zamorani, G.  1994b, \aap\ 289, L34

\bibitem[Elbaz, Arnaud \& Boehringer 1995]{a2163} \reference{} Elbaz, D.,
	Arnaud, M., and Boehringer, H.  1995, \aa\ 293, 337

\bibitem[Ellingson et~al. 1989]{ell89} \reference{} Ellingson, E., Yee, 
	H.~K.~C., Green, R.~F., and Kinman, T.D. 1989, \aj\ 97, 6

\bibitem[Ellingson, Green \& Yee 1991]{egy91} \reference{} Ellingson, E., 
	Green, R.~F., and Yee, H.~K.~C.  1991, \apj\ 378, 476

\bibitem[Ellingson, Yee \& Green 1991]{eyg91} \reference{} Ellingson, E., Yee, 
	H.~K.~C., and Green, R.~F. 1991, \apj\ 371, 36 (EYG91)

\bibitem[Ellingson \& Yee 1994]{ey94} \reference{} Ellingson, E., and Yee, 
     H.~K.~C. 1994, \apjs\ 92, 33


\bibitem[Elvis, Lockman \& Wilkes 1989]{elw89} \reference{} Elvis, M., Lockman,
	F.J., and Wilkes, B.J. 1989, \aj\ 97, 777

\bibitem[Evans 1996]{eva96} \reference{} Evans, A.S.  1996, PhD Thesis,
	University of Hawaii.

\bibitem[Evrard 1990]{evr90} \reference{} Evrard, A.E.  1990, \apj\ 363, 349

\bibitem[Fabian \& Crawford 1990]{fc90} \reference{} Fabian, A.~C., and 
	Crawford, C.~S. 1990, \mnras\ 247, 439	

\bibitem[Fabian \& Crawford 1995]{fc95} \reference{} Fabian, A.~C., and 
	Crawford, C.~S. 1995, \mnras\ 274, L63	

\bibitem[Fabian et~al. 1986]{fab86} \reference{} Fabian, A.~C., Arnaud, K.~A.,
     	Nulsen, P.~E.~J., and Mushotzky, R.~F. 1986, \apj\ 305, 9

\bibitem[Fabian 1992]{fab92} \reference{} Fabian, A.~C., in {\it Clusters
	and Superclusters of Galaxies}, ed. A.C. Fabian (Dordrecht: Kluwer), 151

\bibitem[Fabian 1994]{fab94} \reference{} Fabian, A.~C., \araa\ 32, 277

\bibitem[Fabian et~al. 1994]{fea94} \reference{} Fabian, A.~C., Shioya, Y.,
	Iwasawa, K., Nandra, K., Crawford, C., Johnstone, R., Kuneida, H.,
	McMahon, R., Makishima, K., Murayama, T., Ohashi, T., Tanaka, Y.,
	Taniguchi, Y., and Terashima, Y.  1994, \apjl\ 436, L54

\bibitem[Fabianno et~al. 1984]{fab84} \reference{} Fabbiano, G., Miller, L,
	Trinchieri, G., Longair, M., and Elvis, M.  1984, \apj\ 277, 115

\bibitem[Fabricant, McClintock \& Bautz 1991]{fmb91} \reference{} Fabricant,
	D.G., McClintock, J.E., and Bautz, M.W.  1991, \apj\ 381, 33

\bibitem[Fabricant, Bautz \& McClintock 1994]{fbm94} \reference{} Fabricant,
	D.G., Bautz, M.W., and McClintock, J.E.  1994, \aj\ 107, 8


\bibitem[Fruscione et~al. 1995]{fru95} \reference{} Fruscione, A., Drake, J.J.,
	McDonald, K., and Malina, R.F. 1995, \apj\ 441, 726



\bibitem[Gioia \& Luppino 1994]{gl94} \reference{} Gioia, I.M., and Luppino, 
     G.A. 1994, \apjs\ 94, 583

\bibitem[Golombek, Miley \& Neugebauer 1988]{gmn88} \reference{} Golombek, D.,
	Miley, G.K., and Neugebauer, G.  1988, \aj\ 95, 26 




\bibitem[Hall et~al. 1995]{paper1} \reference{} Hall, P.~B., Ellingson, E., 
	Green, R.~F., and Yee, H.~K.~C.  1995, \aj, 110, 513.

\bibitem[Harris 1984]{h84} \reference{} Harris, D. E., ed., 1984, The \ein\
	Observatory Revised Users Manual (Cambridge, Harvard/CfA)

\bibitem[Heckman et~al. 1989]{heck89} \reference{} Heckman, T.M., Baum, S.A.,
	van Breugel, W.J.M., \& McCarthy, P.  1989, ApJ 338, 48

\bibitem[Henry et~al. 1992]{hen92} \reference{} Henry, J.P., Gioia, I.M.,
     	Maccacaro, T., Morris, S.L., Stocke, J.T., and Wolter, A.  1992,
	 \apj\ 386, 408

\bibitem[Henry \& Henriksen 1986]{hh86} \reference{} Henry, J.P., and 
	Henriksen, M. J. 1986, \apj\ 301, 689 (HH86)

\bibitem[Hill \& Lilly 1991]{hl91} \reference{} Hill, G.~J., and Lilly, S.~J.
	1991, \apj\ 367, 1

\bibitem[Hines \& Wills 1993]{hw93} \reference{} Hines, D.~C., and Wills, B.~J.
	1993, \apj\ 415, 82

\bibitem[Hughes, Birkinshaw \& Huchra 1995]{hbh95} \reference{}  Hughes, J.P.,
	Birkinshaw, M., and Huchra, J.P.  1995, \apjl\ 448, L93

\bibitem[Hutchings, Crampton \& Campbell 1984]{hut84} \reference{} Hutchings, 
	J.~B., Crampton, D., and Campbell, B.  1984, \apj\ 280, 41 

\bibitem[Hutchings 1987]{hut87} \reference{} Hutchings, J.~B. 1987, \apj\ 320,
	122

\bibitem[Hutchings \& Neff 1991a]{hn91a} \reference{} Hutchings, J.~B., and 
	Neff, S.~G.  1991, \aj\ 101, 434

\bibitem[Hutchings \& Neff 1991b]{hn91b} \reference{} Hutchings, J.~B., and 
	Neff, S.~G.  1991, \aj\ 101, 2001

\bibitem[Hutchings et~al. 1996]{hut96} \reference{} Hutchings, J.~B., Gower,
	A.~C., Ryneveld, S., and Dewey, A.  1996, \aj\ 111, 2167

\bibitem[Jackson et~al. 1996]{j96} \reference{} Jackson, N., Tadhunter, C., 
	Sparks, W.B., Miley, G.K., and Macchetto, F.  1996, \aap\ 307, L29

\bibitem[Johnstone et~al. 1992]{joh92} \reference{} Johnstone, R.M., Fabian,
	A.C., Edge, A.C., and Thomas, P.A.  1992, \mnras\ 255, 431

\bibitem[Jones \& Forman 1984]{jf84} \reference{} Jones, C., and Forman, W.,
     1984, \apj\ 276, 38

\bibitem[Kii et~al. 1991]{kii91} \reference{} Kii, T., Williams, O.R., Ohashi,
	T., Awaki, H., Hayashida, K., Inoue, H., Kondo, H., Koyama, K., Makino,
	F., Makishima, K., Saxton, R.D., Stewart, G.C., Takano, S., Tanaka, Y.,
	and Turner, M.J.L.  1991, \apj\ 367, 455

\bibitem[Kleinmann et~al. 1988]{kle88} \reference{} Kleinmann, S.G., Hamilton,
	D., Keel, W.C., Wynn-Williams, C.G., Eales, S.A., Becklin, E.E., and
	Kuntz, K.D. 1988, \apj\ 328, 161

\bibitem[Kolman et~al. 1991]{kol91} \reference{} Kolman, M., Halpern, J.P.,
	Shrader, C.R., and Filippenko, A.V.  1993, \apj\ 373, 57

\bibitem[Kolman et~al. 1993]{kol93} \reference{} Kolman, M., Halpern, J.P.,
	Shrader, C.R., Filippenko, A.V., Fink, H.H., and Schaeidt, S.G. 1993,
	\apj\ 402, 514


\bibitem[Kriss et al. 1996]{kri96} \reference{} Kriss, G., Krolik, J., Grimes,
	J., Tsvetanov, Z., Espey, B., Zheng, W., and Davidsen, A.  1996, in
	{\it Emission Lines in Active Galaxies: New Methods and Techniques},
	eds. B.M. Peterson, F.-Z. Cheng, and A.S. Wilson (ASP: San Francisco)
	(astro-ph/9607154)

\bibitem[Lacy, Rawlings \& Hill 1992]{lrh92} \reference{} Lacy, M., Rawlings,
	S., and Hill, G.~J. 1992, \mnras\ 258, 828


\bibitem[Leahy, Muxlow \& Stephens 1989]{lms89} \reference{} Leahy, J.P.,
	Muxlow, T.W.B., and Stephens, P.W.  1989, \mnras\ 239, 401

\bibitem[Le Brun, Bergeron \& Boiss\'e 1996]{lbb96} \reference{} Le Brun, V.,
	Bergeron, J., and Boiss\'e, P.  1995, \aa\ 306, 691

\bibitem[Ledlow \& Owen 1995]{lo95} \reference{} 
	Ledlow, M.~J., and Owen, F~.N.  1995, \aj\ 109, 853

\bibitem[Lee et~al. 1993]{lee93} \reference{} Lee, G., Kriss, G.A., Zheng, W.,
	and Davidsen, A.F.  1993, BAAS 182, 792

\bibitem[Lockman \& Savage 1995]{ls95} \reference{} Lockman, F.J., and Savage, 
     B.D. 1995, \apjs\ 97, 1 (LS95)

\bibitem[Longair \& Seldner 1979]{ls79} \reference{} Longair, M.S. and Seldner,
	M.  1979, \mnras\ 189, 433

\bibitem[Luppino \& Gioia 1995]{lg95} \reference{} Luppino, G.A., and Gioia,
	I.M.  1995, \apjl\ 445, L77


\bibitem[Mathieu \& Spinrad 1981]{ms81} \reference{} Mathieu, R.D., and 
	Spinrad, H.  1981, \apj\ 251, 485

%
%

\bibitem[McNamara et~al. 1996]{mcn96} \reference{} McNamara, B.R., Jannunzi,
	B.T., Elston, R., Sarazin, C., and Wise, M.  1996, \apj, in press

\bibitem[Miley \& Hartsuijker 1978]{mh78} \reference{} Miley, G.K., and
     Hartsuijker, A.P. 1978, \aaps\ 34, 129 

\bibitem[Morse 1994]{mor94} \reference{} Morse, J. A. 1994, \pasp\ 106, 675 

\bibitem[Navarro, Frenk \& White 1996]{nfw96} \reference{}  Navarro, J.F.,
	Frenk, C.S., and White, S.D.M.  1996, \apj\ 462, 563

\bibitem[Nesci, Perola \& Wolter 1994]{npw94} \reference{} Nesci, R.,
	Perola, G.C., and Wolter, A.  1994, \aaps\ 299, 34

\bibitem[Neugebauer et~al. 1986]{neg86} \reference{} Neugebauer, G., Soifer,
 	B.~T., Miley, G.~K.; Clegg, P.~E  1986, \apjl\ 308, L1

\bibitem[Nichol et~al. 1994]{nic94} \reference{} Nichol, R.C., Ulmer, M.P., 
     Kron, R.G., Wirth, G.D., Koo, D.C.  1994, \apj\ 432, 464

\bibitem[Nulsen, Stewart \& Fabian 1984]{nsf84} \reference{} Nulsen, P.E.J.,
	Stewart, G.C., and Fabian, A.C.  1984, \mnras\ 208, 185


\bibitem[O'Dea et~al. 1996]{ode96} \reference{} O'Dea, C.P., Worrall, D.M.,
	Baum, S.A., and Stanghellini, C.  1996, \aj\ 111, 92



\bibitem[Padovani \& Urry 1991]{pu91} \reference{} Padovani, P., \& Urry, C.M.
	1991, \apj\ 368, 373

\bibitem[Papadopoulos et~al. 1995]{pap95} \reference{} Papadopoulos, P. P.,
	Seaquist, E.R., Wrobel, J.M., and Binette, L.  1995, \apj\ 446, 150


\bibitem[Perley \& Taylor 1991]{pt91} \reference{} Perley, R.A., and Taylor, 
	G.B.  1991, \aj\ 101, 1623


\bibitem[Pierre et~al. 1994a]{pie94a} \reference{} Pierre, M., Boehringer, H.,
	Ebeling, H., Voges, W., Schuecker, P., Cruddace, R., and MacGillivray,
	H.  1994, \aa\ 290, 725

\bibitem[Pierre et~al. 1994b]{pie94b} \reference{} Pierre, M., Hunstead, R.,
	and Unewisee, A.  1994, in {\it Cosmological Aspects of X-Ray Clusters
	of Galaxies}, ed. W.C. Seiter, (Dordrecht:Kluwer), p. 73


\bibitem[Pravdo \& Marshall 1984]{pm84} \reference{} Pravdo, S.H., and Marshall,
	F.E.  1984, \apj\ 281, 570

\bibitem[Prestage \& Peacock 1988]{pp88} \reference{} Prestage, R.M., and 
	Peacock, J.A.  1988, \mnras\ 230, 131

\bibitem[Quinento \& Cersosimo 1993]{qc93} \reference{} Quinento, Z.M.,
	and Cersosimo, J.C.  1993, A\&AS 97, 435

\bibitem[Rawlings 1994]{raw94} \reference{} Rawlings, S.  1994, in
	{\it The Physics of Active Galactic Nuclei}, eds. G.V. Bicknell,
	M.A. Dopita, and P.J. Quinn (San Franciso: ASP), 253

\bibitem[Raymond \& Smith 1977]{rs77} \reference{} Raymond, J.C., and Smith,
      B.W.  1977, \apjs\ 35, 419

\bibitem[Rector, Stocke \& Ellingson 1995]{rse95} \reference{} Rector, T.A.,
	Stocke, J.T., and Ellingson, E.  1995, \aj\ 110, 1492

\bibitem[Rees et~al. 1982]{ree82} \reference{} Rees, M.J., Begelman, M.C.,
	Blandford, R.D., and Phinney, E.S.  1982, Nature 295, 17

\bibitem[Reynolds \& Fabian 1996]{rf96} \reference{} Reynolds, C.S., and
	Fabian, A.C.  1996, \mnras\ 278, 479

\bibitem[Richstone, Loeb \& Turner 1992]{rlt92} \reference{} Richstone, D., 
	Loeb, A., and Turner, E.L.  1992, \apj 393, 477

\bibitem[Roche et~al. 1995]{roc95} \reference{} Roche, N., Shanks, T., Almaini,
	O., Boyle, B.J., Georgantopoulos, I., Stewart, G.C., and Griffiths, R.E.
	1995, \mnras\ 276, 706


\bibitem[Sarazin 1988]{sar88} \reference{} Sarazin, C.~L.  1988, {\it X-Ray
	Emission from Clusters of Galaxies}, (Cambridge: Cambridge) 


\bibitem[Saxton et~al. 1993]{sax93} \reference{} Saxton, R.D., Turner, M.J.L.,
	Williams, O.R., Stewart, G.C., Ohashi, T., and Kii, T.  1993, 
	\mnras\ 262, 63

\bibitem[Schindler et~al. 1996]{sch96} \reference{} Schindler, S., Hattori, M.,
   Neumann, D.M., and Boehringer, H.  1996, \aaps, submitted (astro-ph/9603037)

\bibitem[Schmidt \& Green 1986]{sg86} \reference{} Schmidt, M., and Green, R.F.
	1986, \apj\ 305, 68

\bibitem[Schneider et~al. 1992]{sch92} \reference{} Schneider, D.P., Bahcall,
	J.N., Gunn, J.E., and Dressler, A.  1992, \aj\ 103, 1047

\bibitem[Schwartz et~al. 1991]{sch91} \reference{} Schwartz, D.A., Bradt, H.V.,
	Remillard, R.A., and Tuohy, I.R.  1991, \apj\ 376, 424

\bibitem[Scoville et~al. 1991]{sco91} \reference{} Scoville, N.Z., Sargent, 
	A.I., Sanders, D.B., and Soifer, B.T.  1991, \apjl\ 366, L5

\bibitem[Smail \& Dickinson 1995]{sd95} \reference{} Smail, I., and Dickinson,
	M.  1995, \apjl\ 455, 99

\bibitem[Smail et~al. 1995]{sma95} \reference{} Smail, I., Couch, W.J., Ellis,	
	R.S., and Sharples, R.M.  1995, \apj\ 440, 501

\bibitem[Smail et~al. 1997]{sma97} \reference{} Smail, I., Ellis, R.S., 
	Dressler, A., Couch, W.J., Oemler, Jr., A., Sharples, R.M., and
	Butcher, H.  1995, \apj, in press

\bibitem[Smith \& Heckman 1989]{sh89} \reference{} Smith, E.P., and Heckman, 
	T.M.  1989, \apjs\ 64, 365

\bibitem[Soifer et~al. 1996]{soi96} \reference{} Soifer, B.T., Neugebauer, G.,
	Armus, L., and Shupe, D.L.  1996, \aj\ 111, 649

\bibitem[Soker \& Sarazin 1988]{ss88} \reference{} Soker, N., and Sarazin, C.L.
	1988, \apj\ 327, 66


\bibitem[Spinrad \& Stauffer 1982]{ss82} \reference{} Spinrad, H., and Stauffer,
	J.R.  1982, \mnras\ 200, 153

\bibitem[Stocke \& Perrenod 1981]{sp81} \reference{} Stocke, J.~T. and Perrenod,
     S.~C. 1981, \apj\ 245, 375

\bibitem[Stocke et~al. 1992]{sto92} \reference{} Stocke, J.~T., Morris, S.~L.,
	Weymann, R.~J. and Foltz, C.~B.  1992, \apj\ 396, 487

\bibitem[Stockton, Ridgway \& Lilly 1994]{srl94} \reference{}
	Stockton, A., Ridgway, S.E., and Lilly, S.J.  1994, \aj\ 108, 414

\bibitem[Struble \& Rood 1987]{sr87} \reference{} Struble, M.F., and Rood, 
	H.J. 1987, \apjs\ 63, 543

\bibitem[Teague, Carter \& Gray 1990]{tcg90} \reference{} Teague, P.F.,
	Carter, D., and Gray, P.M.  1990, \aj\ 72, 715


\bibitem[Tsai \& Buote 1996]{tb96} \reference{} Tsai, J.C., and Buote, D.A.
	1996, \mnras\ in press (astro-ph/9510057)

\bibitem[Turner \& Pounds 1989]{tp89} \reference{} Turner, M.J.L., and Pounds,
	K.A.  1989, \mnras\ 240, 833

\bibitem[Ulrich et~al. 1992]{ul92} \reference{} Ulrich, M.-H., Fink, H.H.,
	Schaeidt, S., Baganoff, F., Malkan, M.A., Heidt, J., and Wagner, S.
	1992, \aa\ 266, 183

\bibitem[Vall\'ee \& Bridle 1982]{vb82} \reference{} Vall\'ee, J.P., and
	Bridle, A.H.  1982, \apj\ 253, 479

\bibitem[Wan \& Daly 1996]{wd96} \reference{} Wan, L., and Daly, R.A.  1996,
	\apj, 467, 145


\bibitem[Warwick, Barstow \& Yaqoob 1989]{wby89} \reference{} Warwick, R.S.,
	Barstow, M.A., and Yaqoob, T.  1989, \mnras\ 238, 917

\bibitem[Wellman, Daly \& Wan 1996a]{wdw96a} \reference{} Wellman, G.F., Daly,
	R.A., and Wan, L.  1996a, \apj, in press

\bibitem[Wellman, Daly \& Wan 1996b]{wdw96b} \reference{} Wellman, G.F., Daly,
	R.A., and Wan, L.  1996b, \apj, in press


\bibitem[White et~al. 1994]{a478} \reference{} White, D.A., Fabian, A.C.,
	Allen, S.W., Edge, A.C., Crawford, C.S., Johnstone, R.M., Stewart,
	G.C., and Voges, W.  1994, \mnras\ 269, 589

\bibitem[Wilkes et~al. 1994]{wil94} \reference{} Wilkes, B.J., Tananbaum, H.,
      Worrall, D.M., Avni, Y., Oey, M.S., and Flanagan, J.  1994, \apjs\ 92, 53

\bibitem[Williams et~al. 1992]{wil92} \reference{} Williams, O.R., Turner,
	M.J.L., Steward, G.C., Saxton, R.D., Ohashi, T., Makishima, K., Kii, 
	T., Inoue, H., Makino, F., Hayashida, K., and Koyama, K.  1992,
	\apj\ 389, 157


\bibitem[Worrall, Birkinshaw \& Cameron 1995]{wbc95} \reference{} Worrall,
	D.M., Birkinshaw, M., and Cameron, R.A.  1995, \apj\ 449, 93

%

\bibitem[Yamashita et~al. 1994]{yam94} \reference{} Yamashita, A., Kii, Tsuneo,
	Tashiro, M., Makishima, K., and Ohashi, T.  1994, in {\it New Horizon
	of X-ray Astronomy: First Results from ASCA}, eds. F. Makino and T. 
	Ohashi, (Universal Academy Press:Tokyo), 599

\bibitem[Yaqoob et~al. 1993]{yaq93} \reference{} Yaqoob, T., Serlemitsos, P.J.,
	Mushotsky, R.F., Weaver, K.A., Marshall, F.E., and Petre, R. 1993,
	\apj\ 418, 638

\bibitem[Yaqoob \& Serlemitsos 1994]{ys94} \reference{} Yaqoob, T., and
	Serlemitsos, P.J.  1994, in {\it New Horizon of X-ray Astronomy: First
	Results from ASCA}, eds. F. Makino and T. Ohashi, (Universal Academy
	Press:Tokyo), 337

\bibitem[Yates, Miller \& Peacock 1989]{ymp89} \reference{} Yates, M.G.,
	Miller, L., and Peacock, J.A.  1989, \mnras,  240, 129

\bibitem[Yee \& Ellingson 1993]{ye93} \reference{} Yee, H.~K.~C., and Ellingson,
     E. 1993, \apj\ 411, 43

\bibitem[Yee \& Green 1987]{yg87} \reference{} Yee, H.~K.~C. and Green, R.~F. 
     1987, \apj\ 319, 28


\bibitem[Zabludoff, Huchra \& Geller 1990]{zhg90} \reference{} Zabludoff, 
	A.I., Huchra, J.P., and Geller, M.J.  1990, \apjs\ 74, 1

\bibitem[Zhao et~al. 1993]{3c317} \reference{} Zhao, J.-H., Sumi, D.M., Burns,
	J.O., and Duric, N.  1993, \apj\ 416, 51

\bibitem[Zombeck et~al. 1990]{zom90} \reference{} Zombeck, M.V., Conroy, M.,
     Harnden, F.R., Roy, A., Braeuninger, H., Burkert, W., Hasinger, G., and 
     Predehl, P. 1990, in Proc. SPIE Conference on EUV, X-Ray, and Gamma-Ray
     Instrumentation for Astronomy, ed. O.H.W. Siegmund and H.S. Hudson,
     Proc. SPIE 1344, p. 267

\end{thebibliography}
\end{document}